\documentclass[10pt]{article}

\usepackage{fullpage}
\usepackage{booktabs} 
\usepackage{array}
\usepackage{listings}
\usepackage{algorithm, algpseudocode}
\usepackage{float}
\usepackage{subcaption}
\usepackage{color,soul}
\usepackage{multirow, makecell}
\usepackage{acronym} 
\usepackage{hhline}
\usepackage{enumitem}
\usepackage{comment}
\usepackage{amsmath}
\usepackage{amssymb}
\usepackage{url}
\usepackage{hyperref}
\usepackage{xurl}
\usepackage{times}
\usepackage{multicol}
\usepackage{adjustbox}
\usepackage{tikz}
\usepackage{stfloats}
\usepackage{standalone}
\usepackage{natbib}
\usepackage{longtable}
\usetikzlibrary{trees, positioning, shapes, shapes.geometric, arrows.meta, calc, backgrounds}

\usepackage{lipsum} 
\usepackage{xargs} 
\usepackage{xcolor}
\usepackage[colorinlistoftodos,prependcaption,textsize=tiny]{todonotes}

\newcommandx{\unsure}[2][1=]{\todo[linecolor=red,backgroundcolor=red!25,bordercolor=red,#1]{#2}}
\newcommandx{\change}[2][1=]{\todo[linecolor=blue,backgroundcolor=blue!25,bordercolor=blue,#1]{#2}}

\newcommandx{\info}[2][1=]{\todo[linecolor=OliveGreen,backgroundcolor=width=13cm,height=4cmOliveGreen!25,bordercolor=OliveGreen,#1]{#2}}
\newcommandx{\improvement}[2][1=]{\todo[linecolor=Plum,backgroundcolor=Plum!25,bordercolor=Plum,#1]{#2}}
\newcommandx{\thiswillnotshow}[2][1=]{\todo[disable,#1]{#2}}

\algnewcommand{\LineComment}[1]{\State \(\triangleright\) #1}

\makeatletter
 {\vskip 5pt\begin{list}{}{%
  \setlength{\topsep}{0pt}\setlength{\partopsep}{0pt}\setlength{\parskip}{0pt}%
  \setlength{\parsep}{0pt}\setlength{\labelwidth}{0pt}%
  \setlength{\rightmargin}{0pt}\setlength{\leftmargin}{0pt}%
  \setlength{\labelsep}{0pt}%
  \obeylines\@vobeyspaces\normalfont\ttfamily%
  \item[]}}
 {\end{list}\vskip5pt\noindent}
\makeatother

\begin{document}

\date{February 2026}

\title{Filtered Approximate Nearest Neighbor Search in Vector Databases: System Design and Performance Analysis}

\author{
Abylay Amanbayev, Brian Tsan, Tri Dang, Florin Rusu\\
\{amanbayev, btsan, tridang, frusu\}@ucmerced.edu\\
	University of California Merced
}

\maketitle


\begin{abstract}

Retrieval-Augmented Generation (RAG) applications increasingly rely on Filtered Approximate Nearest Neighbor Search (FANNS) to combine semantic retrieval with metadata constraints. While algorithmic innovations for FANNS have been proposed, there remains a lack of understanding regarding how generic filtering strategies perform within Vector Databases. In this work, we systematize the taxonomy of filtering strategies and evaluate their integration into FAISS, Milvus, and pgvector. To provide a robust benchmarking framework, we introduce a new relational dataset, \textit{MoReVec}, consisting of two tables, featuring 768-dimensional text embeddings and a rich schema of metadata attributes. We further propose the \textit{Global-Local Selectivity (GLS)} correlation metric to quantify the relationship between filters and query vectors.

Our experiments reveal that algorithmic adaptations within the engine often override raw index performance. Specifically, we find that: (1) \textit{Milvus} achieves superior recall stability through hybrid approximate/exact execution; (2) \textit{pgvector}'s cost-based query optimizer frequently selects suboptimal execution plans, favoring approximate index scans even when exact sequential scans would yield perfect recall at comparable latency; and (3) partition-based indexes (IVFFlat) outperform graph-based indexes (HNSW) for low-selectivity queries. To facilitate this analysis, we extend the widely-used \textit{ANN-Benchmarks} to support filtered vector search and make it available online. Finally, we synthesize our findings into a set of practical guidelines for selecting index types and configuring query optimizers for hybrid search workloads.

\end{abstract}

\section{INTRODUCTION}\label{sec:intro}

Vector Databases have emerged as an important component in the modern AI stack, serving as the long-term memory for Large Language Models (LLMs) \cite{jing2024_llm_vecdb}. In particular, they mitigate hallucinations by providing relevant context from external knowledge bases to Retrieval-Augmented Generation (RAG) applications \cite{ji2023_hallucination, lewis2021_rag}. This is done by retrieval from unstructured data, such as text, images, and audio, stored as vector embeddings, using \textbf{Approximate Nearest Neighbor Search (ANNS)} algorithms. While standard ANNS is a well-studied problem with efficient algorithms \cite{malkov2018_hnsw, filtered-diskann, parlay_ann2024, lsh_survey2021} and established benchmarks  \cite{annbenchmarks, aumüller2018_annbenchmarks, vecdbbenchmark}, real-world deployment reveals that semantic similarity alone is rarely sufficient for context retrieval.

\subsection{Motivation}

As vector search systems mature from prototypes to production, they are expected to handle \textit{hybrid queries}, where irrelevant search results are excluded by filters over structured metadata \cite{analyticDBV-hybrid, jd_platform}. This requirement is ubiquitous across domains: a medical RAG system must find patients with similar symptoms (\textit{vector}) who are also within a specific age range (\textit{filter}); a legal search engine must locate relevant cases (\textit{vector}) within a specific jurisdiction (\textit{filter}); and e-commerce platforms must recommend visually similar products (\textit{vector}) that are highly rated (\textit{filter}) \cite{wang2022_navigable_pg, covington2016_youtube_recs}. 

Despite the universality of these queries, the interaction between structured filtering and vector index traversal remains a critical blind spot in current research. While recent algorithmic work has proposed specialized fusion methods (e.g., ACORN \cite{patel_acorn2024}, Filtered-DiskANN \cite{filtered-diskann}, HQANN \cite{wu2022_hqann}), these approaches often sacrifice flexibility, requiring the index to be tuned for specific attributes or workloads. In contrast, general-purpose Vector Databases aim to be schema-agnostic and data-agnostic, necessitating the use of generic strategies like \textit{pre-filtering} and \textit{post-filtering}.

This creates a gap between algorithmic theory and real-world applications. First, the literature lacks a unified taxonomy to describe these strategies, leading to ambiguous definitions and inconsistent implementations across systems. Second, while generic strategies are flexible, their performance limitations and interplay with system components remain underexplored. 

Furthermore, the community lacks the tooling to properly benchmark these scenarios. Existing benchmarks fail to stress these systems realistically; they are predominantly "flat," relying on image-based datasets (like SIFT or GIST) or pure text corpora (like BEIR \cite{thakur2021_beir}) with synthetic or limited metadata. Crucially, they lack robust correlation metrics to distinguish whether a performance drop is due to the system's architecture or the intrinsic "hardness" of the filter-vector distribution. Without a rigorous evaluation of these factors, it remains unclear how architectural choices of Vector Databases impact the trade-off between recall and query latency in filtered workloads.

\subsection{Contributions}

To address these challenges, we present a comprehensive evaluation of generic filtering strategies within vector database systems. Our specific contributions are as follows:

\begin{itemize}
    \item \textbf{Taxonomy and System Analysis.} We systematize the taxonomy of filtering strategies (Pre-, Post-, and Runtime-filtering) and analyze their implementation in three widely-used systems: \textit{FAISS}, \textit{Milvus}, and \textit{pgvector}. We highlight how system-level architectural decisions, such as adaptive query execution and query planning, often override raw algorithmic properties in production environments.

    \item \textbf{Global-Local Selectivity (GLS) Metric.} We propose the \textit{Global-Local Selectivity (GLS)} correlation metric to rigorously quantify the independence between metadata filters and vector neighborhoods. Unlike previous distance-based metrics, GLS normalizes local filter prevalence against global selectivity, providing a robust signal for indexing decisions.

    \item \textbf{FANNS Dataset and Workload.} We conduct our experiments on a new relational dataset, \textit{MoReVec}, comprising two tables: \textit{Movies} and \textit{Reviews}. This dataset features 768-dimensional dense text embeddings and a rich schema of scalar and categorical metadata, enabling the evaluation of filtered ANNS queries. Furthermore, the relational schema is designed to support future benchmarking of ANNS queries with join predicates.

    \item \textbf{Open-source Benchmarking Framework.} We release an extension of the industry-standard \textit{ANN-Benchmarks} framework capable of executing filtered queries with dynamic selectivity targets, facilitating future research into hybrid search performance. Our extended framework, the MoReVec dataset, and GLS correlation analysis tools are publicly available \cite{annbenchmark_extension}.

    \item \textbf{Empirical Insights on System Behavior.} Through extensive experimentation, we reveal counter-intuitive performance characteristics in modern Vector Databases:
    \begin{itemize}
        \item \textit{Algorithmic Adaptation:} \textit{Milvus} achieves superior recall stability at low selectivities compared to standard implementations by employing a hybrid execution model that dynamically switches between ``Dual-Pool'' graph traversal and adaptive brute-force fallbacks.
        \item \textit{Optimizer Limitations:} We identify that \textit{pgvector}'s cost-based optimizer frequently misjudges execution costs, favoring approximate index scans even when exact sequential scans would yield perfect recall at comparable latency.
        \item \textit{Index Capabilities:} Contrary to standard ANNS wisdom, we demonstrate that partition-based indexes (IVFFlat) often outperform graph-based indexes (HNSW) in low-selectivity regimes due to efficient cluster pruning.
        \item \textbf{Practical Guidelines.} We synthesize our findings into a set of actionable guidelines for index selection, parameter tuning, and query plan verification, providing practitioners with a roadmap for optimizing hybrid search workloads in production.
    \end{itemize}

\end{itemize}

\textbf{Organization.} The remainder of this paper is organized as follows. We begin with Section~\ref{sec:related-work}, which reviews recent work in this field, followed by the necessary background on Approximate Nearest Neighbor Search (ANNS) in Section~\ref{sec:preliminaries}. Section~\ref{sec:gls_correlation} describes our proposed Global-Local Selectivity (GLS) correlation metric. Section~\ref{sec:taxonomy} systematizes the taxonomy of filtering strategies and details the specific architectures of the systems evaluated. Section~\ref{sec:benchmark} outlines our benchmarking framework, introducing the MoReVec dataset and workload. Section~\ref{sec:evaluation} presents the experimental results, analyzing the impact of selectivity and system architecture on performance. Finally, Section~\ref{sec:conclusions} concludes with a summary of our findings and potential future directions.

\section{RELATED WORK}\label{sec:related-work}

The systematization of Filtered ANNS (FANNS) draws upon high-dimensional indexing, hybrid search algorithms, and the evolving architecture of modern Vector Databases.

\begin{longtable}{l p{0.78\textwidth}}
\caption{Comparison between this work and other recent FANNS benchmarks.} \label{tab:comparison_summary} \\

\toprule
\textbf{Paper} & \textbf{Description} \\
\midrule
\endfirsthead

\multicolumn{2}{c}%
{{\bfseries \tablename\ \thetable{} -- continued from previous page}} \\
\toprule
\textbf{Paper} & \textbf{Description} \\
\midrule
\endhead

\midrule
\multicolumn{2}{r}{{Continued on next page}} \\
\bottomrule
\endfoot

\bottomrule
\endlastfoot

\multicolumn{2}{l}{\textit{Research Goal \& Framing}} \\* 
\midrule
\textbf{This Work} & 
\textbf{System-Level Analysis}: Focuses on production Vector Databases (FAISS, Milvus, pgvector). Systematizes generic filtering (Pre/Post/Runtime) and investigates how architectural choices (algorithmic adaptability, query planning) override raw algorithmic performance. \\
\textit{Shi et al.}~\cite{fann_bench2025_china_fudan} & 
\textbf{Unified Benchmark}: Focuses on robustness and fairness. Establishes a standardized tuning protocol and classification of strategy families (Filter-then, Search-then, Hybrid) to mitigate implementation bias. \\
\textit{Li et al.}~\cite{fann_bench2025_li} & 
\textbf{Component Taxonomy}: Focuses on internal index levers. Analyzes performance variability through specific components like pruning heuristics, entry-point selection, and edge-filter costs. \\
\textit{Iff et al.}~\cite{iff_fann_bench2025_swiss} & 
\textbf{Algorithm Survey}: Focuses on specialized/fusion methods. Benchmarks state-of-the-art algorithms (ACORN, Filtered-DiskANN) specifically on Transformer-based embeddings. \\

\midrule

\multicolumn{2}{l}{\textit{Dataset Model}} \\*
\midrule
\textbf{This Work} & 
\textbf{Relational (IMDb)}: Two tables (Movies, Reviews) with 768d text embeddings and several real-world attributes. Has 1-to-many foreign key between tables, can be used for ANNS with JOIN-filtering. \\
\textit{Shi et al.}~\cite{fann_bench2025_china_fudan} & 
\textbf{Real-World Curated}: Selection of diverse datasets, unified into a single harness to test general robustness across varying data distributions. \\
\textit{Li et al.}~\cite{fann_bench2025_li} & 
\textbf{Synthetic + Real}: 4 datasets (up to 10M vectors) mixing real and synthetic attributes to allow precise control over selectivity sweeps (0.1\%--100\%). \\
\textit{Iff et al.}~\cite{iff_fann_bench2025_swiss} & 
\textbf{Flat (ArXiv)}: Single table with 2.7M vectors and 11 real-world attributes. Large-scale text data designed to test fusion algorithms. \\

\midrule

\multicolumn{2}{l}{\textit{Key Insights}} \\*
\midrule
\textbf{This Work} & 
\textbf{Architecture Dominates}: Milvus's algorithmic adaptability and brute-force fallbacks stabilize recall; pgvector's optimizer often chooses suboptimal approximate scans. Introduces \textbf{GLS Correlation} to quantify filter independence. \\
\textit{Shi et al.}~\cite{fann_bench2025_china_fudan} & 
\textbf{Tuning Sensitivity}: Standardized tuning significantly alters the ranking of strategies; "Filter-then" vs. "Search-then" trade-offs are highly sensitive to implementation details. \\
\textit{Li et al.}~\cite{fann_bench2025_li} & 
\textbf{Pruning Drivers}: Performance is driven by pruning granularity and edge-traversal costs; provides practical tuning guidelines per attribute type. \\
\textit{Iff et al.}~\cite{iff_fann_bench2025_swiss} & 
\textbf{Scale Failures}: Fusion methods (like ACORN) show promise but can struggle with index size and build times at scale; no universal winner across all selectivities. \\
\end{longtable}

\subsection{Vector Database Architectures}
The integration of ANNS into database systems has followed two paths: \textbf{Specialized Vector Databases} and \textbf{Relational Extensions}. Systems like \textbf{Milvus} \cite{wang_milvus_vecdb} and \textbf{Pinecone} \cite{pinecone} utilize distributed architectures optimized for horizontal scaling and data segmentation. In contrast, extensions like \textbf{pgvector} \cite{pgvector} and \textbf{AnalyticDB-V} \cite{analyticDBV-hybrid} embed vector types into the relational core, leveraging existing cost-based optimizers (CBO). Alternatively, other approaches within the PostgreSQL ecosystem explore decoupling the vector search engine to bypass the overhead inherent from page-based design \cite{jin2026fast}. A critical research gap exists in understanding how these architectural choices—rather than just the underlying algorithms—impact performance. Recent studies comparing RDBMS and specialized Vector DBs \cite{yunan2024_rdbms_vs_vecdb} suggest that the performance gap is narrowing, yet the specific failure modes of query optimizers in the presence of high-dimensional vectors are not well-documented.

\subsection{Filtered-ANN Algorithms}
Recent algorithmic work has attempted to integrate metadata constraints directly into the index structure to avoid the ``recall cliff'' associated with post-filtering. \textbf{Filtered-DiskANN} \cite{filtered-diskann} introduced a filter-aware graph construction strategy, while \textbf{HQANN} \cite{wu2022_hqann} utilized joint quantization of vectors and metadata. More recently, \textbf{ACORN} \cite{patel_acorn2024} proposed using hierarchical Navigable Small World graphs to maintain connectivity even under low selectivity predicates. While these ``fusion'' methods offer high efficiency, they often require the index to be specialized for specific attributes, limiting their utility in general-purpose databases that must support ad-hoc queries over arbitrary schemas.

\subsection{ANNS Benchmarking}
The foundations of ANNS are built upon four algorithmic pillars: \textbf{graph-based} (e.g., HNSW \cite{malkov2018_hnsw}, Vamana \cite{subramanya_NEURIPS2019_diskann}), \textbf{partition-based} (e.g., IVFFlat, LSH \cite{lsh_survey2021}), \textbf{quantization-based} (e.g., PQ \cite{jegou2011_pq}), and \textbf{tree-based} methods (e.g., Annoy \cite{annoy}). The performance of these algorithms is typically evaluated via the \textit{ANN-Benchmarks} framework \cite{aumüller2018_annbenchmarks}, which formalizes the recall-vs-throughput trade-off. However, traditional benchmarks focus almost exclusively on ``flat'' vector search, utilizing datasets like SIFT and GIST that lack structured metadata. Consequently, the impact of selection predicates remains under-explored in standard literature.

\subsection{Differentiation from Recent FANNS Benchmarks}
Recent studies have begun to address the complexities of Filtered ANNS. Notably, \textit{Shi et al.}~\cite{fann_bench2025_china_fudan}, \textit{Li et al.}~\cite{fann_bench2025_li} and \textit{Iff et al.}~\cite{iff_fann_bench2025_swiss} provide valuable benchmarks for hybrid vector search. However, our work diverges from these studies in three critical aspects, which are detailed in Table~\ref{tab:comparison_summary}.

\textbf{System Architecture vs. Algorithmic Fusion.} While recent benchmarks primarily evaluate specialized algorithms, including fusion-based approaches (e.g., ACORN, UNG) that tightly couple metadata with index structures, we focus on the \textit{system-level integration} of generic filtering strategies (Pre-, Post-, and Runtime-filtering). We contend that fusion methods, while theoretically efficient, lack the schema-agnosticism required for general-purpose Vector Databases. Our results demonstrate that system architectural choices, such as \textit{Milvus}'s algorithmic adaptability or \textit{pgvector}'s query optimizer cost models, often override raw algorithmic performance.

\textbf{Relational vs. Flat Data Models.} Unlike prior benchmarks that rely on "flat" datasets with simple tag-based metadata \cite{iff_fann_bench2025_swiss, fann_bench2025_li}, we introduce a relational dataset with \textit{Movies} and \textit{Reviews} tables connected via a foreign key, enabling the evaluation of complex query plans (e.g., joins) inherent to real-world database workloads.

\textbf{Query-Filter Relationship Study.} We propose the \textit{Global-Local Selectivity (GLS)} correlation, a normalized metric that explicitly isolates whether a filter enriches or depletes the semantic neighborhood, providing a clearer signal for index selection. To the best of our knowledge, prior benchmarks have not explicitly quantified this correlation between filter predicates and query vector neighborhoods.

\section{PRELIMINARIES}\label{sec:preliminaries}

In this section, we outline the fundamental concepts of vector similarity search, the indexing structures evaluated in this work, and the existing metrics for quantifying filter-query correlations.

\subsection{Vector Representations and Embeddings}

Modern information retrieval relies heavily on \textit{dense vector embeddings} \cite{karpukhin2020-dense} — fixed-dimensional representations of unstructured data (text, images, audio) where semantic similarity is preserved as geometric proximity \cite{mikolov2013_word2vec}. Unlike traditional sparse methods (e.g., BM25 \cite{robertson1994_bm25}, TF-IDF), dense embeddings capture semantic nuance, enabling the retrieval of conceptually related items that may not share exact keywords.

The current state-of-the-art for text embeddings utilizes Transformer-based architectures (e.g., GPT \cite{openai2023gpt4, gpt3}, BERT \cite{devlin2019_bert}) and contrastive learning \cite{li2023_gte_embedding, zhang2025_qwen3_embedding}. These models map variable-length text into a fixed-size vector $v \in \mathbb{R}^d$ (commonly $d=768$ or $d=1024$).

\textbf{MTEB Leaderboard.} The \textit{Massive Text Embedding Benchmark (MTEB)}~\cite{mteb_leaderboard, muennighoff2023_mteb} has emerged as the standard for evaluating embedding quality across diverse tasks, including retrieval \cite{thakur2021_beir}, clustering, and classification. It ranks models based on their generalized performance, guiding the selection of embedding models for production systems.

\subsection{Similarity Metrics}

Vector similarity metrics quantify the semantic relationship between a query vector $q$ and a database vector $v$. These metrics generally fall into two categories: \textit{distance-based metrics}, which measure the geometric distance in vector space, and \textit{angle-based metrics}, which measure directional alignment. While recent works have proposed fusion distances that aggregate vector similarity with metadata distinctness~\cite{wu2022_hqann}, generic Vector Databases predominantly rely on three standard metrics:

\textit{$L_p$-norm}. This metric generalizes the standard Euclidean ($L_2$) and Manhattan ($L_1$) distances. It is defined as $d(v, q) = {||v - q||}_p = (\sum_{i=1}^n |v_i - q_i|^p)^{1/p}$. Euclidean distance ($L_2$) is the most ubiquitous metric. However, it is susceptible to the ``\textit{Curse of Dimensionality}''. As dimensionality increases, the volume of the space expands exponentially, causing the ratio of distances between the nearest and farthest points to approach 1. This loss of contrast can make distance-based discrimination difficult in extremely high-dimensional sparse spaces, though modern dense embeddings (e.g., 768 dimensions) generally retain sufficient structural locality.

\textit{Inner Product}. Defined as $d(v, q) = \langle v, q\rangle = \sum_{i=1}^n v_i q_i$, this metric projects $v$ onto $q$ scaled by the magnitude of $q$. It is computationally efficient but sensitive to vector magnitude; for example, two identical large vectors will yield a higher similarity score than two identical small vectors, which may not be desirable for all semantic tasks.

\textit{Cosine Similarity}. This metric measures the cosine of the angle between two vectors, effectively normalizing the inner product: $d(v, q) = \frac {\langle v, q\rangle} {||v|| ||q||}$. This is mathematically equivalent to the Inner Product of $L_2$-normalized vectors \cite{aggarwal_metric_curse}. It is widely favored in text retrieval as it captures semantic orientation independent of document length (vector magnitude); consequently, it serves as the standard metric for the majority of embedding models evaluated on the MTEB leaderboard \cite{muennighoff2023_mteb}.

\begin{table}[htbp]
\centering
\small
\caption{Notation and key parameters.}
\label{tab:notation}
\begin{tabular}{l l}
\toprule
\textbf{Symbol} & \textbf{Description} \\
\midrule
\multicolumn{2}{l}{\textit{Dataset and Query}} \\
$\mathcal{D}, N$ & Dataset of vectors and its cardinality ($|\mathcal{D}| = N$) \\
$\mathcal{Q}$ & Set of query vectors \\
$d$ & Vector dimensionality (768 in our experiments) \\
$k$ & Number of nearest neighbors to retrieve \\
\midrule
\multicolumn{2}{l}{\textit{Filtering and Selectivity}} \\
$\phi(v)$ & Filter predicate function, returns 1 if valid, 0 otherwise \\
$\sigma_g$ & \textbf{Global Selectivity}: Fraction of $\mathcal{D}$ satisfying $\phi$ \\
$\sigma_l$ & \textbf{Local Selectivity}: Fraction of $k$-NN satisfying $\phi$ \\
$r$ & \textbf{Selectivity Ratio}: $\sigma_l / \sigma_g$ (as proposed in \cite{xia2026_fannsQO}) \\
$\rho_q$ & \textbf{GLS Correlation}: Per-query filter-vector independence \\
$\bar{\rho}$ & Mean GLS correlation across $\mathcal{Q}$ \\
\midrule
\multicolumn{2}{l}{\textit{Index Parameters}} \\
$M$ & HNSW: Maximum number of edges per node \\
$ef_{const}$ & HNSW: Candidate queue size during construction \\
$ef_{search}$ & HNSW: Candidate queue size during search \\
$n_{probe}$ & IVFFlat: Number of clusters (centroids) to search \\
\bottomrule
\end{tabular}
\end{table}

\subsection{Indexing and Search Quality}

The $k$-NN problem involves scanning through the entire set of vectors while maintaining a record of the $k$-nearest neighbors found thus far. This exhaustive search requires the distance calculation between all $N$ vectors with the query vector, resulting in a time complexity of $O(Nd)$ for a dataset of $N$ vectors. Although polynomial, this approach demands substantial computational resources, especially for high-workload systems with large vector collections. To address that, ANNS algorithms have been developed to trade small amounts of accuracy for significant speed improvements \cite{li2019_ann_tkde}.

A wide variety of these indexing strategies exists, supported by established evaluation frameworks \cite{annbenchmarks}. They are usually classified in four categories: \textit{graph-based}, such as HNSW, Vamana, \textit{partition-based} such as IVF, Locality Sensitive Hashing, \textit{quantization-based}, such as PQ, and \textit{tree-based}, such as ANNOY, k-d tree algorithms \cite{subramanya_NEURIPS2019_diskann, jegou2011_pq, lsh_survey2021, bentley1975_kdtree, muja2014_flann}. Some algorithms may combine multiple approaches into one. Below we will discuss the details of some commonly implemented ANNS algorithms: Inverted File Index, which is a partition-based algorithm, and HNSW which is graph-based algorithm.

The \textbf{Inverted File Index} (IVF) is a data structure used in information retrieval systems to locate documents containing specific terms efficiently. It establishes a link between the terms $t_i$ and the set of documents $D_i$ containing them. For the ANNS, Inverted file indexes are often complemented by Voronoi diagrams or k-means clustering to partition a space into regions based on proximity to a set of points or objects \cite{jegou2011_pq, sivic2003_videogoogle}. By associating each vector with a point in the Voronoi diagram or cluster centroid, queries can efficiently pinpoint relevant documents, thereby optimizing the retrieval process. This leverages spatial relationships between documents and also decreases the search space for ANNS problems. When the association is based on pure distance (L2 or Cosine) this algorithm is known as \textit{IVFFlat}.

\textbf{Hierarchical Navigable Small-Worlds (HNSW)} is a graph-based indexing algorithm that extends the Navigable Small World (NSW) model to enable logarithmic scalability for high-dimensional vector search \cite{malkov2018_hnsw}. The standard NSW algorithm iteratively builds a graph by connecting each vector to its nearest neighbors and adding some long-range links that connect different regions of the vector space.

The core idea of HNSW is to separate connections by length: long-range links (shortcuts) are placed in upper layers to facilitate rapid traversal across the vector space, while short-range links remain in the lower layers for fine-grained local search. For that, vectors are inserted consecutively into the graph structure. For every inserted element, an integer maximum layer $l$ is randomly selected with an exponentially decaying probability. This probabilistic assignment ensures that while all vectors are present in the lowest layer, only a sparse subset of "hub" vertices populates the upper layers, acting as expressways into the graph.

The search process begins at the highest layer and proceeds top-down. The algorithm greedily traverses the graph elements in the current layer until a nearest neighbor is reached, at which point it descends to the next layer below. This "zoom-in" strategy allows the algorithm to ignore the vast majority of distance computations. Upon reaching the base layer, a candidate priority queue is maintained to track the nearest neighbors found. The size of this queue is controlled by the parameter $ef_{search}$, which serves as a tunable trade-off between search latency and recall quality.

\textbf{Recall@k} serves as the primary metric for evaluating the accuracy of the approximate search. It measures the proportion of the true nearest neighbors that are successfully retrieved by the index and defined as: 

\begin{equation} \text{Recall}@k = \frac{\text{TP}}{\text{TP} + \text{FN}} = \frac{\text{TP}}{k} \end{equation}. 

Here, \textit{True Positives} (TP) represent the number of relevant, while \textit{False Negatives} (FN) represent the true neighbors missed by the approximation. Since $\text{TP} + \text{FN} = k$, this simplifies to the fraction of the top-$k$ ground truth found.

There is also another important characteristic, which is the speed of query processing, which is measured as number of queries processed per second or \textbf{QPS} (Queries Per Second). Indexing methods are usually evaluated as a graph where QPS is a function of Recall \cite{aumüller2018_annbenchmarks}.

\subsection{Filter Types}

Filtering predicates in vector search generally fall into two primary categories based on the nature of the constraint: metadata-based and distance-based.

\textbf{Metadata-based filters} enforce constraints on the structured attributes associated with vector embeddings (e.g., tags, timestamps, or unique identifiers). These are the direct equivalent of the \texttt{WHERE} clause in relational SQL and are now standard across modern Vector Databases \cite{milvus, pinecone, pgvector, qdrant}. They can be subdivided into: 
\begin{itemize} 
    \item \textit{Categorical filters}: Exact matches on discrete labels (e.g., \texttt{category = 'electronics'}) or set membership checks (e.g., \texttt{tags CONTAINS 'urgent'}). There is a benchmark dedicated specifically for such filters with many algorithms aimed to solve it \cite{simhadri2024_results_bigann, parlay_ann2024, jin2026_curator_filter}.
    \item \textit{Scalar filters}: Numerical range queries on continuous values (e.g., \texttt{price > 50.0} or \texttt{year BETWEEN 2020 AND 2024}). 
    \item \textit{String filters}: Pattern matching operations using standard substring search (e.g., \texttt{LIKE}) or regular expressions. 
\end{itemize}

\textbf{Distance-based filters}, often referred to as a separate problem, known as \textit{Range Search}, leverage the distance between vectors to prune results. Instead of a fixed $k$-nearest neighbors retrieval, these filters return all vectors falling within a specified radius of the query (e.g., \texttt{cosine similarity > 0.8}) \cite{chavez2001_range_search}.

A third, emerging paradigm is Semantic filtering, which operates outside these traditional categories. Here, constraints are defined by natural language predicates rather than structured schema fields. These systems typically employ a secondary LLM step to evaluate conditions that require semantic reasoning (e.g., ``find papers associated with \textit{vector databases}'' where the topic is inferred rather than explicitly tagged) \cite{patel_lotuslang2024}.

In this work, we focus exclusively on \textbf{Metadata-based filters}—specifically scalar inequalities—as they represent the standard integration point for Hybrid Search in production Vector Databases (see Section \ref{ssec:queries}).

Throughout this work, we adopt the standard database definition where selectivity $\sigma$ represents the \textit{fraction} of valid rows satisfying filter predicates. Consequently, we use the term \textit{low selectivity} to denote a strict filter (where $\sigma \to 0$, e.g., $1\%$), and \textit{high selectivity} to denote a relaxed filter (where $\sigma \to 1$, e.g., $50\%$).

\subsection{Query-Filter Correlation: Distance-based Approach}

In FANNS, the integration of filters with vector similarity introduces challenges in balancing recall and latency. These challenges are amplified when filter attributes exhibit a correlation (or anti-correlation) with the spatial distribution of the vector embeddings.

State-of-the-art approaches, such as ACORN~\cite{patel_acorn2024}, quantify this relationship using \textit{distance-based approach}. In particular, they compare the distance distribution of filtered results against a random baseline, defining the query-filter correlation $C(\mathcal{D}, \mathcal{Q})$ for a workload $\mathcal{Q}$ over dataset $\mathcal{D}$ as the expected difference in minimum distances between the query vector and the equally-sized filtered and random subsets of $\mathcal{D}$:

\begin{equation}
C(\mathcal{D}, \mathcal{Q}) = \mathbb{E}_{(x_i, p_i) \in \mathcal{Q}} \left[ \mathbb{E}_{R_i} [g(x_i, R_i)] - g(x_i, X_{p_i}) \right],
\end{equation}

where:
\begin{itemize}
\item $X_{p_i}$ is the subset of vectors satisfying the filter predicate $p_i$.
\item $R_i$ is a random subset drawn uniformly from $\mathcal{D}$, with $|R_i| = |X_{p_i}|$.
\item $g(x, S) = \min_{y \in S} \text{dist}(x, y)$ is the distance from query $x$ to the nearest vector in set $S$.
\end{itemize}

Intuitively, a positive $C(\mathcal{D}, \mathcal{Q})$ indicates that the filtered vectors are spatially clustered around the query (closer than random chance), suggesting high local density. Conversely, a negative value implies the filtered subset is sparser or more distant than a random sample. While this metric effectively captures geometric ``hardness,'' it is susceptible to the influence of the \emph{local neighborhood density}, the varying average distance from each query $q$ to its neighbors. By averaging minimum distances, that metric conflates global clustering with per-query locality, failing to distinguish enrichment (high local selectivity) from mere proximity in dense regions. Furthermore, estimating such a correlation metric for query optimization remains an open challenge.

\section{GLOBAL-LOCAL SELECTIVITY CORRELATION}\label{sec:gls_correlation}

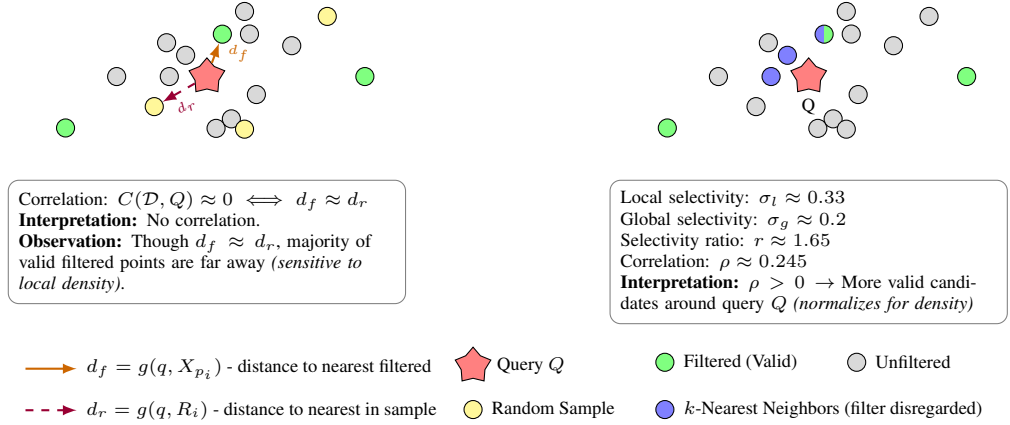
\begin{figure*}[t] 
\centering
\begin{tikzpicture}[
    node distance=2.5cm,
    point/.style={circle, fill=gray!30, draw=black, inner sep=0pt, minimum size=2.4mm, font=\small},
    filtered/.style={circle, fill=green!50, draw=black, inner sep=0pt, minimum size=2.4mm, font=\small},
    sampled/.style={circle, fill=yellow!50, draw=black, inner sep=0pt, minimum size=2.4mm, font=\small},
    topk/.style={circle, fill=blue!50, draw=black, inner sep=0pt, minimum size=2.4mm, font=\small},
    topk_filt/.style={circle, draw=black, inner sep=0pt, minimum size=2.4mm, font=\small, path picture={
            \fill[blue!50] (path picture bounding box.south west)
                rectangle (path picture bounding box.north -| path picture bounding box.center);
            \fill[green!50] (path picture bounding box.south east)
                rectangle (path picture bounding box.north east -| path picture bounding box.center);}
        },
    query/.style={star, fill=red!50, draw=black, minimum size=1.5mm, font=\small},
    boundary/.style={circle, draw=blue, dashed, thick, fill=blue!5, fill opacity=0.1},
    arrow_dist/.style={-latex, thick, color=orange!80!black},
    arrow_rand/.style={-latex, dashed, thick, color=purple!80!black},
    textbox/.style={rectangle, draw=black!50, fill=white, rounded corners, align=left, font=\scriptsize, inner sep=4pt}
]

    \begin{scope}[local bounding box=PanelA]
        \node[anchor=south, font=\small\bfseries] at (0, 1.8) {a) Distance-Based Correlation (e.g., ACORN)};
        
        \node (q1) [query, label=right:{\scriptsize}] at (0,0) {};
        
        \foreach \ang/\rad in {20/1.2, 45/0.8, 60/1.0, 135/0.4, 140/0.7, 180/0.5, 180/1.2, 280/0.7, 300/0.65, 340/0.7}
            \node[point] at (\ang:\rad) {};
            
        \node (f1) [filtered] at (70:0.6) {}; 
        \node [filtered] at (200:2.0) {};
        \node [filtered] at (0:2.1) {};
        
        \node (rand) [sampled] at (-0.7, -0.4) {};
        \node (rand2) [sampled] at (1.6, 0.8) {};
        \node (rand3) [sampled] at (0.5, -0.7) {};

        \draw[arrow_dist] (q1) -- node[right, font=\tiny, xshift=1pt] {$d_f$} (f1);
        \draw[arrow_rand] (q1) -- node[below, sloped, font=\tiny] {$d_r$} (rand);
        
        \node[textbox, below=1.2cm of q1, text width=5cm] {
            Correlation: $C(\mathcal{D}, Q) \approx 0 \iff d_{f} \approx d_{r}$ \\
            \textbf{Interpretation:} No correlation.\\
            \textbf{Observation:} Though $d_{f} \approx d_{r}$, majority of valid filtered points are far away \textit{(sensitive to local density)}.
        };
    \end{scope}

    \begin{scope}[shift={(8.0,0)}, local bounding box=PanelB]
        \node[anchor=south, font=\small\bfseries] at (0, 1.8) {b) Global-Local Selectivity (GLS) Correlation};        
        \node (q2) [query, label=below:{\scriptsize Q}] at (0,0) {};
        
        \foreach \ang/\rad in {20/1.2, 45/0.8, 60/1.0, 140/0.7, 180/1.2, 280/0.7, 300/0.65, 340/0.7}
            \node[point] at (\ang:\rad) {};
        \node [filtered] at (200:2.0) {};
        \node [filtered] at (0:2.1) {};
        \node [topk] at (135:0.4) {};
        \node [topk] at (180:0.5) {};
        \node [topk_filt] at (70:0.6) {};
        \node [point] at (-0.7, -0.4) {};
        \node [point] at (1.6, 0.8) {};
        \node [point] at (0.5, -0.7) {};
        
        \node[textbox, below=1.2cm of q2, text width=5cm] {
            Local selectivity: $\sigma_l \approx 0.33$ \\
            Global selectivity: $\sigma_g \approx 0.2$\\
            Selectivity ratio: $r \approx 1.65$ \\ Correlation: $\rho \approx 0.245$ \\
            \textbf{Interpretation:} $\rho > 0 \to$ More valid candidates around query $Q$ \textit{(normalizes for density)} \\ 
        };
    \end{scope}

    \node[anchor=north, yshift=-0.3cm] at ($(PanelA.south)!0.5!(PanelB.south)$) {
        \begin{tikzpicture}[scale=0.85]
            \node[query, label=right:{\scriptsize Query $Q$}] at (0, 0.6) {};
            \node[sampled, label=right:{\scriptsize Random Sample}] at (0, -0.24) {};
            \node[filtered, label=right:{\scriptsize Filtered (Valid)}] at (3.0, 0.5) {};
            \node[topk, label=right:{\scriptsize $k$-Nearest Neighbors (filter disregarded)}] at (3.0, -0.24) {};
            \node[point, label=right:{\scriptsize Unfiltered}] at (6.0, 0.5) {};
            \draw[arrow_dist] (-7,0.25) -- (-6.2,0.25) node[right, black, font=\scriptsize] {$d_f = g(q, X_{p_i})$ - distance to nearest filtered};
            \draw[arrow_rand] (-7,-0.4) -- (-6.2,-0.4) node[right, black, font=\scriptsize] {$d_r = g(q, R_i)$ - distance to nearest in sample};
        \end{tikzpicture}
    };

\end{tikzpicture}
\caption{Comparison between correlation metrics. (a) The distance-based correlation~\cite{patel_acorn2024} is sensitive to geometric density, potentially misidentifying dense clusters as non-correlated. (b) The proposed GLS correlation normalizes for density by comparing local selectivity $\sigma_l$ to the global baseline $\sigma_g$.}
\label{fig:gls_vs_acorn_comparison}
\end{figure*}

We introduce the \textbf{Global-Local Selectivity (GLS) Correlation}, a metric that quantifies the relationship between a filter's global selectivity and its selectivity within the query's local neighborhood. It explicitly reveals whether filters enrich or deplete the relevant neighborhood around the query vector on a per-query level.

Consider a dataset $\mathcal{D}$ comprising $N$ vectors $v \in \mathbb{R}^d$. A filter $\phi: \mathcal{D} \to {0, 1}$ is applied to select vectors satisfying a predicate, with the assumption that the filtered subset is non-empty (i.e., $|{v \in \mathcal{D} \mid \phi(v) = 1}| > 0$; empty filters are ignored in analysis). The \emph{global filter selectivity}, denoted $\sigma_g$, quantifies the baseline prevalence of the filter across the entire dataset:

\begin{equation}
\sigma_g = \frac{|\{v \in \mathcal{D} \mid \phi(v) = 1\}|}{N} \in (0, 1]
\end{equation}
\noindent
This serves as a corpus-level prior, independent of any specific query.

For a set of query vectors $\mathcal{Q}$ where each $q \in \mathcal{Q} \subseteq \mathbb{R}^d$, we define the local neighborhood $\mathcal{N}_q$ as the $k$-nearest neighbors of $q$ in $\mathcal{D}$, with $|\mathcal{N}_q| = k$. The \emph{local filter selectivity} $\sigma_l$ then measures the fraction of this neighborhood that passes the filter:

\begin{equation}
\sigma_l = \frac{|\{v \in \mathcal{N}_q \mid \phi(v) = 1\}|}{k} \in [0, 1]
\end{equation}
\noindent
Here, $\sigma_l > \sigma_g$ indicates local enrichment (filter attributes cluster with similar vectors), while $\sigma_l < \sigma_g$ suggests depletion (anti-correlation).

To derive a directional correlation from these selectivities, we compute the raw \emph{selectivity ratio}:

\begin{equation}
r = \frac{\sigma_l}{\sigma_g} \in [0, \infty),
\end{equation}
\noindent
which normalizes the local prevalence against the global baseline: $r = 1$ denotes neutrality, $r > 1$ enrichment, and $0 < r < 1$ depletion. We note that recent work has independently identified this ratio as a critical factor for estimating the computational cost of filtered ANN search~\cite{xia2026_fannsQO}. However, the unbounded nature of $r$ limits its utility in two key ways. First, it complicates \textbf{aggregation}: a simple mean is highly sensitive to outliers, where rare global filters appearing in local neighborhoods can yield exponentially large ratios that skew the dataset-level statistic. Second, for \textbf{optimization}, cost models typically require normalized inputs (e.g., $[-1, 1]$ or $[0, 1]$) to establish stable thresholds for plan switching. An unbounded metric makes it difficult to define consistent heuristics for when to prefer specific indexing strategies. We thus apply the bilinear (M\"obius) transformation to obtain the per-query \emph{filter-vector selectivity correlation} $\rho_q$:

\begin{equation}
\rho_q = \frac{r - 1}{r + 1} \in [-1, 1)
\end{equation}
\noindent
This maps the ratio symmetrically to $[-1, 1)$, where $\rho_q = 0$ signals neutrality, $\rho_q > 0$ positive correlation (enrichment strength), and $\rho_q < 0$ negative correlation (depletion strength). The transformation is monotonic and differentiable, with derivative $f'(r) = \frac{2}{(r+1)^2} > 0$. The inverse, for denormalization, is $r = (1 + \rho_q)/(1 - \rho_q)$ for $\rho_q \in [-1, 1)$.

Finally, to aggregate across queries and assess systematic entanglement, we define the \emph{mean filter-vector selectivity correlation}

\begin{equation}
\bar{\rho} = \frac{1}{|\mathcal{Q}|} \sum_{q \in \mathcal{Q}} \rho_q \in [-1, 1)
\end{equation}
\noindent

We can also obtain a decent estimate of this metric by finding the selectivity of approximate $k$-nearest neighbors using an ANN-index \textit{(approximate local selectivity)} and the selectivity of a sample drawn uniformly from our dataset \textit{(approximate global selectivity)}.

\section{TAXONOMY OF FILTERING STRATEGIES AND SYSTEMS}\label{sec:taxonomy}

\begin{figure*}[htbp]
\centering
    \begin{tikzpicture}[
        node distance=2.5cm,
        point/.style={circle, fill=gray!30, draw=black, inner sep=0pt, minimum size=3mm, font=\small},
        start/.style={circle, fill=yellow!50, draw=black, inner sep=0pt, minimum size=3mm, font=\small, path picture={
            \fill[yellow!50] (path picture bounding box.south west)
                rectangle (path picture bounding box.north -| path picture bounding box.center);
            \fill[green!50] (path picture bounding box.south east)
                rectangle (path picture bounding box.north east -| path picture bounding box.center);}
        },
        filtered/.style={circle, fill=green!50, draw=black, inner sep=0pt, minimum size=3mm, font=\small},
        gt/.style={circle, fill=blue!50, draw=black, line width=1.5pt, inner sep=0pt, minimum size=3mm, font=\small},
        gt/.style={circle, draw=black, line width=1.5pt, inner sep=0pt, minimum size=3mm, font=\small, path picture={
            \fill[blue!50] (path picture bounding box.south west)
                rectangle (path picture bounding box.north -| path picture bounding box.center);
            \fill[green!50] (path picture bounding box.south east)
                rectangle (path picture bounding box.north east -| path picture bounding box.center);}
        },
        query/.style={star, fill=red!60, draw=black, minimum size=3mm, font=\small},
        edge/.style={gray, thin},
        candidate_edge/.style={green, ultra thick, -Stealth},
        path_edge/.style={blue, ultra thick, -Stealth},
        maxdcircle/.style={circle, draw=blue, dashed, thin, opacity=0.8, inner sep=0pt, minimum size=3mm},
        draw_circle/.style={circle, draw=blue, thick, inner sep=0pt, minimum size=3mm}
    ]
    
        \node (q) [query, label=above:Q] at (-0.9, -0.3) {};
        \node (p1) [gt, label=above:P1] at (-1.2, 1.05) {};
        \node (p2) [start, label=above:P2] at (1.8, 1.8) {};
        \node (p3) [point, label=above:P3] at (0.6, 1.8) {};
        \node (p4) [filtered, label=right:P4] at (2.4, -0.6) {};
        \node (p5) [gt, label=below:P5] at (0.1, 0.6) {};
        \node (p6) [point, label=left:P6] at (-2.2, 1.2) {};
        \node (p7) [point, label=right:P7] at (1, -1.5) {};
        \node (p8) [gt, label=left:P8] at (-1.5, -1.2) {};
        \node (p9) [filtered, label=above:P9] at (-0.6, 2.5) {};
        \node (p10) [filtered, label=below:P10] at (0.9, -0.6) {};
        \node (p11) [point, label=below:P11] at (2.7, 0.6) {};
        \node (p12) [point, label=left:P12] at (-3.2, -0.) {};
    
        \draw [edge] (p1) -- (p5);
        \draw [edge] (p1) -- (p6);
        \draw [edge] (p1) -- (p9);
        \draw [edge] (p2) -- (p4);
        \draw [edge] (p2) -- (p11);
        \draw [edge] (p2) -- (p3);
        \draw [edge] (p3) -- (p9);
        \draw [edge] (p3) -- (p5);
        \draw [edge] (p12) -- (p7);
        \draw [edge] (p4) -- (p10);
        \draw [edge] (p5) -- (p10);
        \draw [edge] (p11) -- (p8);
        \draw [edge] (p7) -- (p10);
        \draw [edge] (p8) -- (p7);
        \draw [edge] (p9) -- (p6);
        \draw [edge] (p6) -- (p12);
        \draw [edge] (p8) -- (p12);
        \draw [edge] (p4) -- (p11);
    
        \draw [path_edge] (p2) -- (p3) node[midway, below, font=\small] {1};
        \draw [candidate_edge] (p2) -- (p11) node[midway, below, font=\tiny] {};
        \draw [candidate_edge] (p2) -- (p4) node[midway, below, font=\tiny] {};
        \draw [path_edge] (p3) -- (p5) node[midway, right, font=\small] {2};
        \draw [candidate_edge] (p3) -- (p9) node[midway, below, font=\tiny] {};
        \draw [candidate_edge] (p5) -- (p10) node[midway, below, font=\tiny] {};
        \draw [path_edge] (p5) -- (p1) node[midway, below, font=\small] {3};
        \draw [candidate_edge] (p1) -- (p6) node[midway, below, font=\tiny, color=red] {};
        \draw [path_edge] (p5) -- (p10) node[midway, right, font=\small] {4};
        \draw [path_edge] (p1) -- (p9) node[midway, right, font=\small] {5};
        \draw [candidate_edge] (p10) -- (p7) node[midway, below, font=\tiny, color=red] {};
        \draw [candidate_edge] (p9) -- (p6) node[midway, below, font=\tiny, color=red] {};
        \draw [candidate_edge] (p10) -- (p4) node[midway, below, font=\tiny, color=red] {};
        
        \draw [maxdcircle] (q) circle (1.4cm);
            
        \draw [blue, thick] (p1) circle (0.25cm);
        \draw [blue, thick] (p5) circle (0.25cm);
        \draw [blue, thick] (p10) circle (0.25cm);
    
        \node[right, xshift=6.5cm, yshift=1cm] at (q.east) {
            \begin{tabular}{ll}
                \tikz\node[start, label=right:Starting point (passing filter)] {}; & \\[ -16pt] \\
                \tikz\node[filtered, label=right:Points passing filter] {}; &  \\[ -16pt] \\
                \tikz\node[point, label=right:Irrelevant points] {}; & \\[ -16pt] \\
                \tikz\node[gt, label=right:Top-3 true results] {}; & \\[ -16pt] \\
                \tikz\node[draw_circle, label=right:Top-3 search results] {}; & \\[ -16pt] \\
                \tikz\node[query, label=right:Query point] {}; & \\[ -16pt] \\
                \tikz\node[maxdcircle, label=right: True result max-distance circle] {}; & \\[-16pt]\\
                \tikz\draw[path_edge] (0,0) -- (1.0,0) node [text=black]{Graph traversal path}; & \\[ -16pt] \\
                \tikz\draw[candidate_edge] (0,0) -- (1.0,0) node[text=black] {Candidate paths}; \\
            \end{tabular}
        };
    \end{tikzpicture}

    \centering
    \small
    \begin{tabular}{c|c|c|c|c|c}
    \hline
    \textbf{Step} & \textbf{Current} & \textbf{Results} & \textbf{Neighbors} & \textbf{Candidates} & \textbf{Visited} \\
    \hline
    0 & P2  & []             & P3, P4, P11 & \textbf{P3}, P4, P11  & P2, P3, P4, P11                          \\
    1 & P3  & P5, P9, P2     & P2, P5, P9  & \textbf{P5}, P9, P4   & P2, P3, P4, P5, P9, P11                  \\
    2 & P5  & P1, P5, P10    & P1, P3, P10 & \textbf{P1}, P10, P9  & P1, P2, P3, P4, P5, P9, P10, P11         \\
    3 & P1  & P1, P5, P10    & P5, P6, P9  & \textbf{P10}, P9 $^*$ & P1, P2, P3, P4, P5, P6, P9, P10, P11     \\
    4 & P10 & P1, P5, P10    & P4, P5, P7  & \textbf{P9} $^*$      & P1, P2, P3, P4, P5, P6, P7, P9, P10, P11 \\
    5 & P9  & P1, P5, P10    & P1, P3, P6  & [] $^*$               & 
    P1, P2, P3, P4, P5, P6, P7, P9, P10, P11 \\ \hline
    \end{tabular}
    
    $^*$ Candidates are not updated since there is no neighbor point $n \in Neighbors$, s.t. $d(n, q) < max(d(Results, q))$ or $|Results| < ef_\text{search}$
    \vspace{-0.2cm}
    \caption{Graph traversal using the \textit{pre-filtering} strategy in 1-layer HNSW index with $m=3$ and search parameters $k=3$, $ef_\text{search}=3$.}
    \vspace{-0.2cm}
    \label{fig:o1_1}

\end{figure*}

\subsection{FANNS Strategies}

When the set of returned results must also satisfy certain metadata (or structured data) constraints, such a query is generally known as a \textit{hybrid query} or \textit{filtered ANNS} or \textit{filtered vector search}. In these cases, we need to apply a selection operation on the subset of elements that meet the given condition, which is commonly referred to as \textit{filtering}. In relational database management systems (RDBMS), this operation is known as the application of a \textit{selection predicate}. In this work, we will use the terms \textit{Selection Predicate} and \textit{Filter} interchangeably, assuming they are equivalent across different systems.

We draw from recent surveys on vector databases~\cite{survey2023vdbms1, survey2023vdbms2} to outline the primary strategies for Filtered ANNS and provide their exact definitions. There are generally three main methods used for filtering in hybrid queries: \textit{pre-filtering}, \textit{runtime-filtering}, and \textit{post-filtering}. While these terms may appear intuitively straightforward, the literature reveals a lack of systematic understanding, resulting in varying definitions and conceptual overlaps across works~\cite{survey2023vdbms2, lin2025_survey_fanns, wang2022_navigable_pg, fann_bench2025_china_fudan}, particularly with respect to pre-filtering. The specifics of each method depend on the underlying index type (e.g., graph-based or partition-based).

We exclude \textit{fusion-based} methods \cite{patel_acorn2024, wu2022_hqann}, where metadata values are fused into either the index construction or the distance calculation. We omit these approaches from this study primarily because they lack schema-agnosticism. By tightly coupling metadata with the vector index, these methods compromise system flexibility and scalability. Fusion methods struggle with sparse or high-dimensional attribute data; if many vectors lack certain attributes or if the total number of attributes is large, fusing all of them into a single, compact vector space or metric becomes a complex, non-trivial problem. Furthermore, fusion methods may require a priori knowledge of the query workload to effectively integrate metadata into the index or distance metric, limiting their applicability to arbitrary schemas or ad-hoc queries. For these reasons, they are not treated here as generic filtering strategies for existing ANNS indexes.

\begin{figure*}[htbp]
\centering
\small
    \begin{tikzpicture}[
        node distance=0.5cm,
        process/.style={rectangle, minimum width=2.5cm, minimum height=0.5cm, text centered, draw=black, fill=gray!30},
        decision/.style={diamond, aspect=2, minimum width=2.5cm, minimum height=0.5cm, text centered, draw=black, fill=red!30, inner sep=0pt},
        io/.style={trapezium, trapezium left angle=70, trapezium right angle=-70, minimum width=2.5cm, minimum height=0.5cm, text centered, draw=black, fill=blue!30},
        startend/.style={rectangle, rounded corners, minimum size=0.5cm, draw=black, fill=green!30},
        arrow/.style={-latex, thick}
    ]
    
        \begin{scope}[xshift=-2.5cm]
            \node[font=\bfseries] (title) {Pre-filtering (Milvus, FAISS):};
            \node (start) [startend, below=0.1cm of title] {Start};
            \node (filter) [process, below=of start] {Filter Metadata $\rightarrow$ Bitset};
            \node (loopstart) [io, below=of filter] {Loop: Traverse Graph};
            \node (check) [process, below=of loopstart] {Check Bitset};
            \node (check2) [process, below=of check] {Update Results};
            \node (loopend) [io, below=of check2] {Check Stopping Criteria};
            \node (finish) [startend, below=of loopend] {Return Top-K};
    
            \tikzset{
                arrow/.style={
                    -{Stealth[scale=1]}, 
                    line width=1pt, 
                    draw=black
                }
            }
            
            \draw [arrow] (start) -- (filter);
            \draw [arrow] (filter) -- (loopstart);
            \draw [arrow] (loopstart) -- (check);
            \draw [arrow] (check) -- (check2);
            \draw [arrow] (check2) -- (loopend);
            \draw [arrow] (loopend.east) -- node[below, pos=0.5, font=\small] {Continue} ([xshift=1.2cm]loopend.east) |- ([xshift=1cm]loopstart.east) -- (loopstart);
            \draw [arrow] (loopend.south) -- node[right, pos=0.5, font=\small] {Stop} (finish.north);
    
        \end{scope}

        \begin{scope}[xshift=3.cm]
            \node[font=\bfseries] (title2) {Runtime-filtering:};
            \node (start2) [startend, below=0.1cm of title2] {Start};
            \node (loopstart2) [io, below=of start2] {Loop: Traverse Graph};
            \node (filter2) [process, below=of loopstart2] {Filter Metadata};
            \node (filter3) [process, below=of filter2] {Update Results};
            \node (loopend2) [io, below=of filter3] {Check Stopping Criteria};
            \node (finish2) [startend, below=of loopend2] {Return Top-K};
    
            \tikzset{
                arrow/.style={
                    -{Stealth[scale=1.]}, 
                    line width=1pt, 
                    draw=black
                }
            }
            
            \draw [arrow] (start2) -- (loopstart2);
            \draw [arrow] (loopstart2) -- (filter2);
            \draw [arrow] (filter2) -- (filter3);
            \draw [arrow] (filter3) -- (loopend2);
            \draw [arrow] (loopend2.east) -- node[below, pos=0.5, font=\small] {Continue} ([xshift=1.2cm]loopend2.east) |- ([xshift=1cm]loopstart2.east) -- (loopstart2);
            \draw [arrow] (loopend2.south) -- node[right, pos=0.5, font=\small] {Stop} (finish2.north);
    
        \end{scope}

        \begin{scope}[xshift=8.5cm]
            \node[font=\bfseries] (title2) {Post-filtering (pgvector):};
            \node (start2) [startend, below=0.1cm of title2] {Start};
            \node (loopstart2) [io, below=of start2] {Loop: Traverse Graph};
            \node (check2) [process, below=of loopstart2] {Update Results};
            \node (loopend2) [io, below=of check2] {Check Stopping Criteria};
            \node (filter2) [process, below=of loopend2] {Filter Metadata};
            \node (filter3) [process, below=of filter2] {Update Results};        
            \node (finish2) [startend, below=of filter3] {Return Top-K};
    
            \tikzset{
                arrow/.style={
                    -{Stealth[scale=1.]}, 
                    line width=1pt, 
                    draw=black
                }
            }
            
            \draw [arrow] (start2) -- (loopstart2);
            \draw [arrow] (loopstart2) -- (check2);
            \draw [arrow] (check2) -- (loopend2);
            \draw [arrow] (loopend2.east) -- node[below, pos=0.5, font=\small] {Continue} ([xshift=1.2cm]loopend2.east) |- ([xshift=1cm]loopstart2.east) -- (loopstart2);
            \draw [arrow] (loopend2.south) -- node[right, pos=0.5, font=\small] {Stop} (filter2);
            \draw [arrow] (filter2) -- (filter3);
            \draw [arrow] (filter3) -- (finish2.north);
    
        \end{scope}
        
    \end{tikzpicture}
    \caption{Filtering strategies for ANNS.}
    \label{fig:o1_0}
\end{figure*}
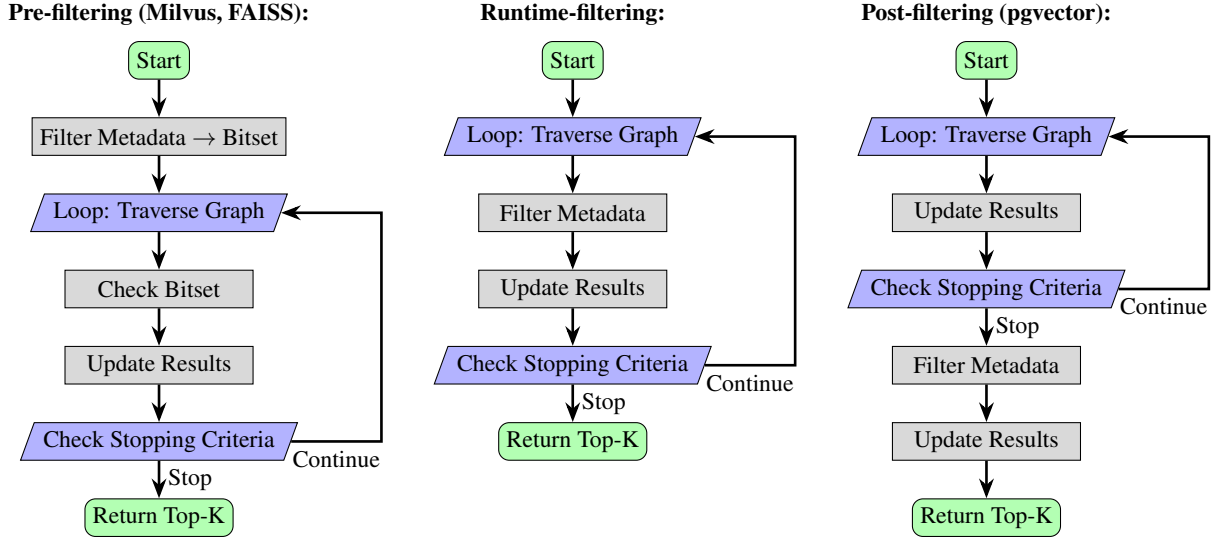

\subsubsection{Pre-filtering + ANNS}

The filter predicate is evaluated upfront to prepare a bitset indicating which vectors pass the filter criteria. This bitset is then used to guide the vector search, ideally reducing the effective search space by masking irrelevant points. However, in graph-based indices, this does not necessarily hold.

For \textit{graph-based indexes}, distances must still be calculated during traversal to ensure the connectivity of the graph and navigate toward the query vector. The bitset is consulted to determine whether a visited vector should be added to the priority queue of candidates.

For \textit{partition-based indexes} (e.g., IVFFlat), the bitset allows checking the filter before computing distances for each vector. This approach prunes the search space early, avoiding unnecessary distance computations on non-qualifying points. 

\textit{Note on \textbf{Pre-filtering + kNNS}.} Sometimes, for queries with low selectivities, vector search engines can apply filter and search exhaustively among survived points. It is important to distinguish it from Pre-filtering + ANNS, as it solves problem exactly rather than approximately.

\subsubsection{Runtime Filtering}

In this approach, attribute retrieval and predicate evaluation occur lazily, triggered only when a candidate vector is accessed during the search. While this on-demand pattern can introduce latency due to random I/O access, it significantly reduces the total count of predicate evaluations. This trade-off is particularly advantageous in memory-constrained environments or when vector filters are computationally expensive, such as complex regular expressions, string pattern matching, or conditions involving joins across multiple relations.

In \textit{graph-based indexes}, attributes are fetched for the vectors traversed during the graph navigation. To mitigate the I/O overhead from these repeated fetches, batching strategies can be employed, such as traversing multiple hops and fetching attributes for all visited points at once. 

In \textit{partition-based indexes}, attributes are fetched only for vectors within the search space (relevant clusters in IVFFlat). This targeted fetching typically incurs no additional I/O overhead beyond the standard cluster access, as the attributes of all points within the target cluster are loaded together.

We exclude Runtime Filtering from our primary evaluation as it is not natively exposed by the standard APIs of the evaluated in-memory systems (FAISS, pgvector). While integral to disk-based systems like DiskANN \cite{filtered-diskann} to mask I/O latency, adapting it for in-memory HNSW requires custom index modification that deviates from the 'off-the-shelf' system behavior this paper seeks to systematize.

\subsubsection{Post-filtering}

The filter is applied after the vector search retrieves a set of candidate vectors. This decouples the filtering from the search but risks including false candidates (similar vectors that fail the filter), which are discarded post-search. It often performs well in terms of search latency but may suffer from reduced recall. This method is commonly implemented in relational DBMS systems, such as pgvector in PostgreSQL, and potentially in proprietary systems like Oracle Database.

In \textit{graph-based indexes}, filtering occurs only on the limited set of vectors in the candidate priority queue (e.g., in HNSW, a larger queue size increases traversal latency). For low selectivity predicates uncorrelated with vector similarities, this can lead to incomplete top-$k$ results (fewer than $k$ survivors).

In \textit{partition-based indexes}, search parameters typically allow probing a larger candidate set (e.g., setting $nprobe$ in IVFFlat to check multiple clusters), reducing the risk of incomplete results compared to graph indexes, though it persists for extreme selectivity.

\subsubsection{Query Optimization for FANNS}

Integration of vector search into relational DBMS introduces additional possibilities to write more complex queries, involving relational operators, such as joins with other tables or filters applied across joined relations. This introduces a fundamental dilemma in query optimization: whether to first do vector search and then execute the "relational" part of query, or to first complete the "relational" part of the query to determine the set of relevant vectors before performing the similarity search.

The query optimizer relies on cardinality estimates to choose between execution plans. However, to prioritize vector search, the optimizer requires accurate \textit{selectivity} estimates to determine the expansion factor—calculating $k / \text{selectivity}$ candidates to ensure $k$ results remain after filtering. However, without reliable estimates, the system is forced to complete the relational join or filter first to generate the subset of qualifying IDs. This subset is then passed as a constraint to the vector engine, ensuring accuracy at the cost of potential latency.

\subsection{Vector Search Systems}

All systems designed for vector similarity search can be divided into three main categories: Vector Extensions for Relational Databases (e.g., PG-Vector \cite{pgvector}, PASE \cite{yang_pase}, VBASE \cite{zhang_vbase}), Specialized Vector Databases (e.g., Milvus \cite{wang_milvus_vecdb}, Pinecone \cite{pinecone}, Weaviate \cite{weaviate}), and Vector Search Libraries (e.g., FAISS \cite{faiss2025_vecdb}, ParlayANN). We will test representative systems from each category.

\begin{table}[htbp]
\centering
\small
\begin{tabular}{lcccc}
\toprule
System & Scalar & Categ. & Str. Match. & Filtering Strategies \\
\midrule
\textbf{Extensions} & & & & \\
PG-Vector & $\checkmark$ & $\checkmark$ & $\checkmark$ & Cost-based (Post / Pre + ANNS) \\
PASE & $\checkmark$ & $\checkmark$ & $\checkmark$ & IVFFlat / HNSW integrated w/ PG \\
VBASE & $\checkmark$ & $\checkmark$ & $\checkmark$ & Relaxed Monotonicity (Kernel-integrated) \\
\midrule
\textbf{Specialized DBs} & & & & \\
Milvus & $\checkmark$ & $\checkmark$ & $\checkmark$ (regex) & Pre-filtering + ANNS \\
Qdrant & $\checkmark$ & $\checkmark$ & $\checkmark$ (exact/substr) & Pre-filtering (Payload Indexing) \\
Weaviate & $\checkmark$ & $\checkmark$ & $\checkmark$ (BM25) & Pre-filtering (Inverted Index) \\
Pinecone & $\checkmark$ & $\checkmark$ & $\checkmark$ (prefix) & Pre-filtering (Metadata Index) \\
Vespa & $\checkmark$ & $\checkmark$ & $\checkmark$ (ling) & Multi-stage (Pre-filter/Rank) \\
Elasticsearch & $\checkmark$ & $\checkmark$ & $\checkmark$ (full-text) & Hybrid (Pre/Post/Rescore) \\
\midrule
\textbf{Libraries} & & & & \\
FAISS (w/ NumPy) & $\checkmark$ & $\times$ & $\times$ & Pre-filtering + ANNS (IDSelector) \\
HNSWlib & $\times$ & $\times$ & $\times$ & Post-filtering (during traversal) \\
ParlayANN & $\times$ & $\times$ & $\times$ & Parallel Graph-based Search \\
\bottomrule
\end{tabular}
\caption{Comparison of vector search systems.}
\label{tab:vecdb_systems}
\end{table}

\subsubsection{Vector Extensions}

\textit{Vector Extensions} integrate vector data types directly into relational database columns, thereby retaining core relational capabilities such as ACID compliance and queries with \texttt{JOIN}, \texttt{GROUP BY}-predicates via SQL syntax. A key benefit is seamless adoption within existing relational infrastructures, obviating the need for data migration. Nonetheless, these extensions may exhibit functional constraints compared to purpose-built solutions. We select \textbf{PG-Vector} \cite{pgvector} for its active maintenance and widespread adoption as a PostgreSQL extension \cite{pgvector4aws}. It supports a comprehensive range of filters, encompassing essentially all predicates available in PostgreSQL (e.g., scalar, categorical and string operations via \texttt{WHERE} clauses). The cost-based model of the query optimizer allows switching between filtering strategies, however it doesn't always choose optimal Filtered ANNS queries as shown in Section \ref{sec:evaluation}.

Other notable candidates include \textbf{VBASE} \cite{zhang_vbase}, a system built on PostgreSQL modifies the database kernel to support \textit{''relaxed monotonicity''} for smarter search termination check, and \textbf{PASE}\cite{yang_pase}, an Alibaba-developed extension on PostgreSQL with IVFFlat and HNSW indexes.

\subsubsection{Specialized Vector Databases}

\textit{Specialized Vector Databases} eschew traditional relational schemas and usually don't support SQL-like syntax. They range from schema-less key-value stores to those enforcing predefined schemas for vector collection and often include native support for embedding generation. Compared to vector extensions, specialized vector databases offer performance advantages in indexing and querying, though these can often be mitigated to insignificant levels \cite{yunan2024_rdbms_vs_vecdb}. We opt for \textbf{Milvus} \cite{wang_milvus_vecdb, milvus} due to its open-source nature and robust maintenance. It supports a wide array of filters, including scalar and categorical predicates, and string matching via regular expressions. \textit{Knowhere}, Milvus's vector search engine, is built on top of FAISS with customized optimizations. Milvus primarily utilizes a pre-filtering + ANNS strategy, and employs a logic-based query optimizer, which is evolving in newer versions.

Other representative systems in this category include:

\begin{itemize}
    \item \textbf{Pinecone} \cite{pinecone}: A proprietary, closed-source cloud database. While implementation details are opaque, it claims to utilize a custom graph-based index that integrates metadata filtering directly into the index structure. This approach aims to prevent the graph disconnectivity issues typically associated with aggressive pre-filtering on standard HNSW graphs.
    \item \textbf{Qdrant} \cite{qdrant}: An open-source engine (Rust) that employs a "segment-based" architecture. It maintains independent data structures for metadata (e.g., HashMaps for keywords, B-Trees for numeric ranges). During filtered search, these metadata indices generate a bitset used as a mask to identify qualifying points during vector retrieval.
    \item \textbf{Weaviate} \cite{weaviate}: An open-source engine (Go) that couples an HNSW vector index with a traditional inverted index (posting lists) for scalar data. For filtered queries, the system first consults the inverted index to retrieve an allow-list of object IDs (or a bitmap), which is then used to filter candidates during the HNSW graph exploration.
    \item \textbf{Vespa} \cite{vespa}: An open-source tensor compute engine. Unlike pure vector stores, Vespa executes queries as tensor operations. It can also automatically switch to exact search to process the queries with low selectivity filters.
    \item \textbf{Elasticsearch} \cite{elasticsearch}: Built on Apache Lucene, it utilizes the Lucene implementation of HNSW. The system assesses filter selectivity to decide between an approximate search (checking filters during graph traversal) or an exact brute-force scan of the filtered document set.
\end{itemize}

\subsubsection{Vector Search Libraries}

\textit{Vector Search Libraries}, such as FAISS \cite{faiss2025_vecdb, faiss}, are lightweight, schema-agnostic tools optimized for vector indexing and retrieval. Unlike full databases, they do not handle persistent storage; instead, they operate on pre-loaded vectors for indexing. Filtered ANNS queries require an external bitset to delineate vectors satisfying the filter criteria. We employ \textbf{FAISS} due to its open-source nature and widespread adoption in the field.

Other notable libraries in this category include:

\begin{itemize}
    \item \textbf{HNSWlib} \cite{malkov2018_hnsw}: A header-only C++ library implementing the HNSW graph algorithm. It serves as the core indexing engine for several vector databases (e.g., Chroma).
    \item \textbf{ScaNN} \cite{guo2020_scann}: Developed by Google, this library introduces \textit{anisotropic vector quantization}. Unlike standard product quantization (PQ) which minimizes Euclidean reconstruction error, ScaNN optimizes a loss function that prioritizes directional accuracy, which is critical for Inner Product search.
    \item \textbf{ParlayANN} \cite{parlay_ann2024}: A research library focusing on parallel algorithms for ANNS, including graph indexes (e.g., HNSW, Vamana). Its architecture is designed to minimize thread contention and ensure deterministic behavior on high-concurrency multi-core systems.
\end{itemize}

\section{BENCHMARKING FRAMEWORK}\label{sec:benchmark}

\subsection{ANN-Benchmarks}

Evaluating ANNS algorithms is notoriously difficult due to varying hardware and implementation nuances. ANN-Benchmarks \cite{aumüller2018_annbenchmarks} addresses this by providing a containerized testing environment for each algorithm, accompanied by a YAML configuration file that defines build instructions, hyperparameters, and indexing procedures. Once validated, these are executed on a fixed set of public datasets to evaluate trade-offs between recall (fraction of true nearest neighbors retrieved) and throughput (queries per second, QPS), typically under resource constraints like single-threaded CPU execution.

The benchmarks employ a curated list of diverse, high-dimensional vector datasets, spanning domains like images and text, including word embeddings like GloVe-100 (angular distance, 100 dimensions) and image features like SIFT-128 (Euclidean distance, 128 dimensions), as well as sparse sets such as Kosarak (Jaccard distance, up to 27,983 dimensions), as summarized in Table \ref{tab:datasets}. Evaluations are conducted on standardized hardware, such as AWS r6i instances, to ensure reproducibility, with performance metrics plotted on logarithmic scales to illustrate the recall-throughput trade-off alongside secondary factors like index build time and memory footprint. This framework has benchmarked over 30 algorithms, including prominent libraries such as FAISS, Milvus, PG-Vector, HNSWlib, and Annoy, with contributions from the community keeping the results up to date.

\subsection{Limitations of Existing Datasets}

\begin{table}[htbp]
\centering
\small
\begin{tabular}{lcccccccc}
\hline
Dataset Name & Dim. & Cardinality & Queries & Distance & Domain & Embedding type & $|V|^*$ & $|A|^{**}$ \\
\hline
SIFT-1M & 128 & 1,000,000 & 10,000 & L2 & Images & Hand-crafted & 1 & 0 \\
GIST-1M & 960 & 1,000,000 & 1,000 & L2 & Images & Hand-crafted & 1 & 0 \\
GLOVE-100 & 100 & 1,183,514 & 10,000 & Cosine & Text & Count-based & 1 & 0 \\
MNIST-784 & 784 & 60,000 & 10,000 & L2 & Images & Raw pixels & 1 & 0 \\
NYTimes-256-ang & 256 & 290,000 & 10,000 & Cosine & Text & Word2Vec embeddings & 1 & 0 \\
Kosarak & 27,983 & 74,962 & 500 & Jaccard & Web & Sparse binary & 1 & 0 \\
LAION-5B$^{***}$ & 768 & $5.85 \times 10^{9}$ & -- & Cosine & Multi-modal & CLIP (contrastive) & 2 & 6 \\
ArXiv$^{***}$ & 768 & 2,700,000 & 10,000 & Cosine & Text & Transformer-based & 1 & 10 \\
Wiki-22-12-en$^{}$ & 768 & 35,200,000 & -- & Cosine & Text & Transformer-based & 1 & 5 \\
YFCC$^{***}$ & 4,096 & 100,000,000 & -- & L2 & Images & Deep CNN & 1 & 5 \\
Youtube audio$^{***}$ & 128 & 6,100,000 & -- & L2 & Audio & Hand-crafted & 1 & 2 \\
Youtube video$^{***}$ & 1,024 & 6,100,000 & -- & L2 & Video & Hand-crafted & 1 & 2 \\
\hline
\end{tabular}
\caption{Datasets used for ANNS benchmarking \cite{anndatasets}.}
$^*$ $|V|$ -- number of vectors per each record. \
$^{**}$ $|A|$ -- number of metadata attributes per each record. \
$^{***}$ Not included in ANN-Benchmarks; added here for comparison with emerging text-focused datasets.
\label{tab:datasets}
\end{table}
 
Despite their rigor, existing ANN datasets suffer from a key shortfall: they are predominantly derived from non-text embeddings (e.g., image or generic word vectors), limiting relevance to modern NLP tasks like dense text retrieval. The LAION-5B dataset, a popular large-scale alternative, mitigates scale issues with billions of image-text pairs but restricts attributes to basic categorical labels (e.g., aesthetics scores, content flags), lacking the continuous or versatile metadata needed for nuanced evaluations.

\textit{ArXiv} dataset is a notable exception, with transformer-based embeddings from over 2.7 million arXiv abstracts alongside 11 structured attributes (e.g., numerical citation counts, categorical domains, and temporal metadata). Similarly, \textit{Wikipedia-22-12-en-embeddings} offers a massive scale of 35 million transformer-based vectors derived from Wikipedia. While it includes valuable structured metadata, it introduces a significant structural complexity: unlike standard 1-to-1 mappings, this dataset assigns an arbitrary number of vectors to each logical entity (article) because a separate vector is generated for each paragraph. This creates a "variable cardinality" problem for entity retrieval, complicating benchmarking efforts that must distinguish between vector-level (paragraph) similarity and item-level (article) relevance.

However, these datasets fall short because they cannot support complex joins. Their limited scope—primarily numerical and categorical fields without deeper relational structures—precludes extensibility for query compositions like temporal-range joins with categorical hierarchies. Moreover, attributes like author counts, category listings, and version histories, though scholarly in nature, hold marginal value for broader retrieval tasks beyond niche academic filtering. For completeness, datasets like \textit{YFCC100M} and the \textit{YouTube audio/video} sets, included here for multimodal comparison, are evidently inadequate: YFCC relies on image-centric deep CNN embeddings, while the YouTube datasets use hand-crafted features and minimal metadata.

\subsection{Dataset Design}

To address these limitations, we introduce \textbf{MoReVec} (Movies and Reviews Vectors), a dataset designed to facilitate optimization over vector embeddings and structured attributes.

Drawing from the rich ecosystem of IMDb data used for construction of Join-Ordering Benchmark (JOB) \cite{JOB_benchmark} for query optimizers evaluation in RDBMS, we merge text embeddings of movie synopsis and reviews with available web-sourced metadata. In particular, we generate 768-dimensional vectors using the \textit{gte-base-en-v1.5} model \cite{li2023_gte_embedding}, a high-performing embedding method from the Massive Text Embedding Benchmark (MTEB) leaderboard, applied to textual fields such as movie synopsis and user reviews. All embeddings are $L_2$-normalized, ensuring that cosine similarity is equivalent to inner product for all evaluated systems.

The resulting datasets adhere to extensible schemas detailed in Table \ref{tab:morevec_schema}, enabling advanced hybrid queries that combine semantic similarity searches with attribute-based filtering and extensible join operations. To enable scalability studies, we instantiate each dataset at three cardinalities:

\begin{itemize}
\item Small: $|Movies| = 9,999$; $|Reviews| = 247,286$;
\item Medium: $|Movies| = 99,560$; $|Reviews| = 1,496,493$;
\item Large: $|Movies| = 551,155$, $|Reviews| = 2,598,267$;
\end{itemize}

\begin{table}[htbp]
\centering
\begin{subtable}[t]{0.42\textwidth}
\centering
\small
\begin{tabular}{p{1.2cm}|p{2.6cm}|p{1.6cm}}
\hline
\textbf{Field} & \textbf{Description} & \textbf{Data Type} \\
\hline \hline
mid & Unique alphanumeric identifier of the movie & \verb|VARCHAR| \\ \hline
title & Movie title & \verb|VARCHAR| \\ \hline
year & Year when movie was released & \verb|INTEGER| \\ \hline
genre & Movie genre & \verb|VARCHAR| \\ \hline
avgrating & Average rating of the movie & \verb|FLOAT| \\ \hline
numvotes & Number of votes for the movie & \verb|INTEGER| \\ \hline
mvector & Movie synopsis embeddings & \verb|VECTOR(768)| \\
\hline
\end{tabular}
\caption{Movies table}
\label{tab:movies_schema}
\end{subtable}
\hfill
\begin{subtable}[t]{0.56\textwidth}
\centering
\small
\begin{tabular}{p{1.4cm}|p{4.4cm}|p{1.6cm}}
\hline
\textbf{Field} & \textbf{Description} & \textbf{Data Type} \\
\hline \hline
rid & Unique alphanumeric identifier of the review & \verb|VARCHAR| \\ \hline
uid & Alphanumeric identifier of a user who left review & \verb|VARCHAR| \\ \hline
mid & Foreign key to mid in movies table & \verb|VARCHAR| \\ \hline
rvector & Vector embedding for the review to a movie & \verb|VECTOR(768)| \\ \hline
movierating & Rating assigned to movie by user & \verb|INTEGER| \\ \hline
totalvotes & Total votes for the review & \verb|INTEGER| \\ \hline
likeshare & Share of positive votes for the review & \verb|FLOAT| \\ \hline
quality & Quality of reviews based on their totalvotes and likeshare & \verb|VARCHAR| \\ 
\hline
\end{tabular}
\caption{Reviews table}
\label{tab:reviews_schema}
\end{subtable}
\caption{Schema of the proposed dataset.}
\label{tab:morevec_schema}
\end{table}

\subsection{Workload}\label{ssec:queries}

\begin{figure}[htbp]
\centering
\includegraphics[width=\textwidth]{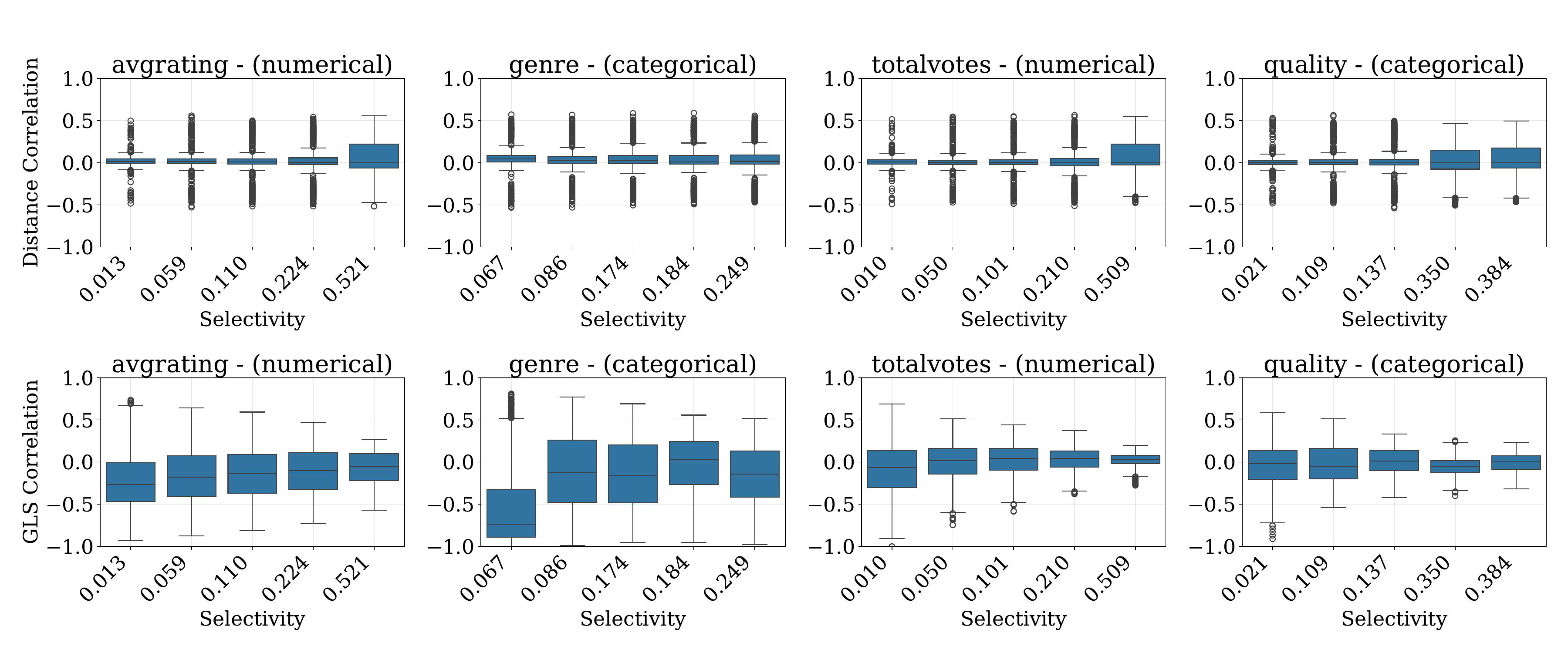} 
\captionof{figure}{Distribution of per-query correlation values $\rho_q$ between metadata attributes and vector embeddings at different target selectivities.}
\label{fig:s4_correlation}
\end{figure}

\textbf{Filters}. Figure~\ref{fig:s4_correlation} shows that the mean \textit{GLS correlation} $\bar{\rho}$ between vectors and metadata attributes tends to 0 for all examined attributes across the breadth of the selectivity range in both datasets. At low selectivity levels ($\sigma_g \leq 0.1$), we observe a notable deviation toward negative values for some attributes. This can be partially attributed to a sampling artifact arising from the finite neighborhood size ($|\mathcal{N}_q| = 2048$); when global selectivity is sufficiently low, the expected number of valid neighbors within the fixed neighborhood is small. Consequently, queries often yield fewer valid local neighbors than the expected value ($\sigma_l<\sigma_g$), driving $\rho_q$ to $-1$ and skewing the mean.

Accounting for this artifact, the results confirm that metadata attributes are mostly independent of the embedding space, allowing us to study the isolated effect of filter selectivity on filtered ANN performance. We therefore apply numeric/scalar filters on the \texttt{avgrating} column (Movies dataset) and \texttt{totalvotes} column (Reviews dataset). Filter thresholds are chosen such that the expected selectivity $\sigma_g$ (fraction of the dataset passing the filter) approximately matches the following target values:

\begin{itemize}
    \item \vspace{-0.5em} Small dataset: $\sigma_g \approx \{0.01; 0.03; 0.05; 0.1; 0.2; 0.5\}$
    \item \vspace{-0.5em} Medium dataset: $\sigma_g \approx \{0.0003; 0.001; 0.01; 0.03; 0.05; 0.1; 0.2\}$
    \item \vspace{-0.5em} Large dataset: $\sigma_g \approx \{0.0003; 0.001; 0.01; 0.03; 0.05; 0.1\}$
\end{itemize}

\begin{figure}[htbp]
\centering
\includegraphics[width=\textwidth]{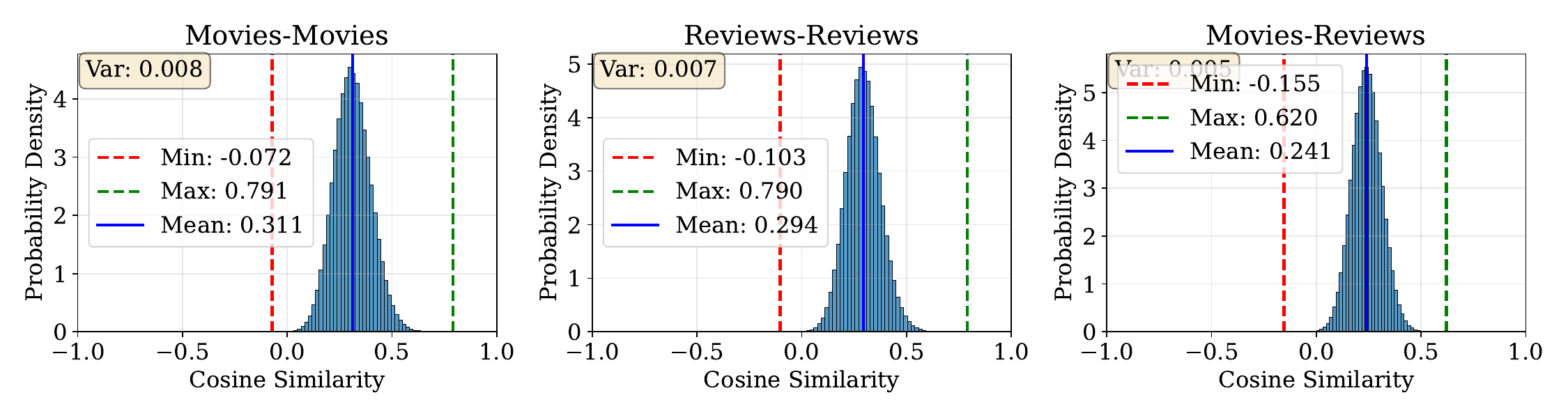}
\vspace{-2.\baselineskip}
\captionof{figure}{Distribution of pairwise cosine distances for 1,000,000 randomly sampled vector pairs. Left: Movie embeddings only; Middle: Review embeddings only; Right: Mixed pairs with one movie and one review embedding each.}
\label{fig:s4_dist_distr}
\end{figure}

\textbf{Queries}. Since semantic-search queries are derived from the same embedding model as the data, we assume that distributions of vectors from \textit{Movies} and \textit{Reviews} tables are similar. This is supported by Figure~\ref{fig:s4_dist_distr}, which shows the distribution of pairwise cosine distances for vector pairs from each table and their mixture.

Accordingly, we randomly sample 1,000 vectors from each dataset to serve as the query set. Each query semantically corresponds to \textit{''find top-}$k$\textit{ closest vectors to query vector }$q$\textit{ where value of scalar attribute is greater than or equal to threshold }$t$\textit{''}. We test four different values of $k$: $k \in \{1, 10, 40, 100\}$. Each of the 1000 query vectors is executed with every filter condition and all four $k$, yielding $1000 \times 7 \times 4 = 28{,}000$ filtered $k$-NN queries per dataset (including no filter runs).

We did not include multi-table \texttt{JOIN} predicates to our query workload. While pgvector supports them syntactically, its query optimizer consistently reverts to sequential scans for such queries, bypassing the vector index entirely. Conversely, schema-agnostic systems like Milvus and FAISS lack the relational operators to execute joins natively.

\section{EMPIRICAL EVALUATION}\label{sec:evaluation}

We perform extensive empirical comparison of FANNS performance across \textit{FAISS}, \textit{Milvus}, and \textit{pgvector}. We focus specifically on the trade-offs between query throughput and recall, as well as the underlying query planning and execution strategies employed by each system. Our main goal is to determine how generic filtering strategies (pre-filtering and post-filtering) perform within real-world Vector Databases, and to what extent system-level architectural choices override the expected algorithmic behavior of the underlying index structures. To this end, our evaluation investigates the following questions:

\begin{itemize}
    \item \textit{How does filter selectivity impact the recall-throughput trade-off?} We examine whether low-selectivity filters universally degrade performance, or whether certain index types and filtering strategies adapt better to sparse result sets.
    
    \item \textit{Does the traditional dominance of HNSW over IVFFlat hold under filtered search?} We test whether the graph-based traversal of HNSW remains superior to the cluster-pruning mechanics of IVFFlat across varying selectivity regimes.
    
    \item \textit{How do system-level architectural decisions affect performance?} We investigate whether factors such as algorithmic adaptability, data segmentation (Milvus) and cost-based query optimization (pgvector) have a greater impact on filtered ANNS performance than the choice of index algorithm itself.
    
    \item \textit{Can query optimizers in relational vector extensions make effective plan selections for hybrid queries?} We analyze whether pgvector's cost-based optimizer reliably selects execution plans that balance recall and latency.
\end{itemize}

To facilitate reproducibility and further research, our extended \textit{Filtered ANN-Benchmarks} framework, the MoReVec relational datasets, and our GLS correlation analysis tools are available on GitHub \cite{annbenchmark_extension}.

\begin{table}[htbp]
\centering
\small
\caption{Software versions used in experiments.}
\label{tab:software_versions}
\begin{tabular}{lll}
\toprule
\textbf{System} & \textbf{Server/Library Version} & \textbf{Notes} \\
\midrule
Milvus & v2.6.6 & pymilvus 2.6.3 \\
FAISS & 1.12.0 (faiss\_hnsw) & CPU version \\
pgvector & 0.8.1 (git main) & PostgreSQL 16 \\
\bottomrule
\end{tabular}
\end{table}

\subsection{Experimental Setup}

\textbf{Indexes.} We evaluate two widely implemented ANN indexing methods: \textit{HNSW} and \textit{IVFFlat}. For the \textbf{HNSW} index, we explore the following hyperparameter configurations:

\begin{itemize}
    \item Construction parameters: $(M, ef_{\text{construction}}) \in \{(5, 25), (10, 50), (15, 75)\}$
    \item Search parameter: $ef_{\text{search}} \in \{1, 10, 40, 100, 200, 500, 1000\}$
\end{itemize}

\noindent For the \textbf{IVFFlat} index, we vary the number of centroids to control the trade-off depending on dataset size:

\begin{itemize}
    \item Construction parameters: $clusters = \{small: (100, 500), medium: (300, 1200), large: (750, 1600)\}$.
    
    These ranges roughly follow the common $C \approx \sqrt{N}$ heuristic, where $C$ is number of clusters, and $N$ is the dataset size.
    
    \item Search parameters: $n_\text{probes} = \{1, 5, 10, 20, 50, 150, 300\}$
\end{itemize}

\begin{figure}[htbp]
\centering
  \includegraphics[width=\textwidth]{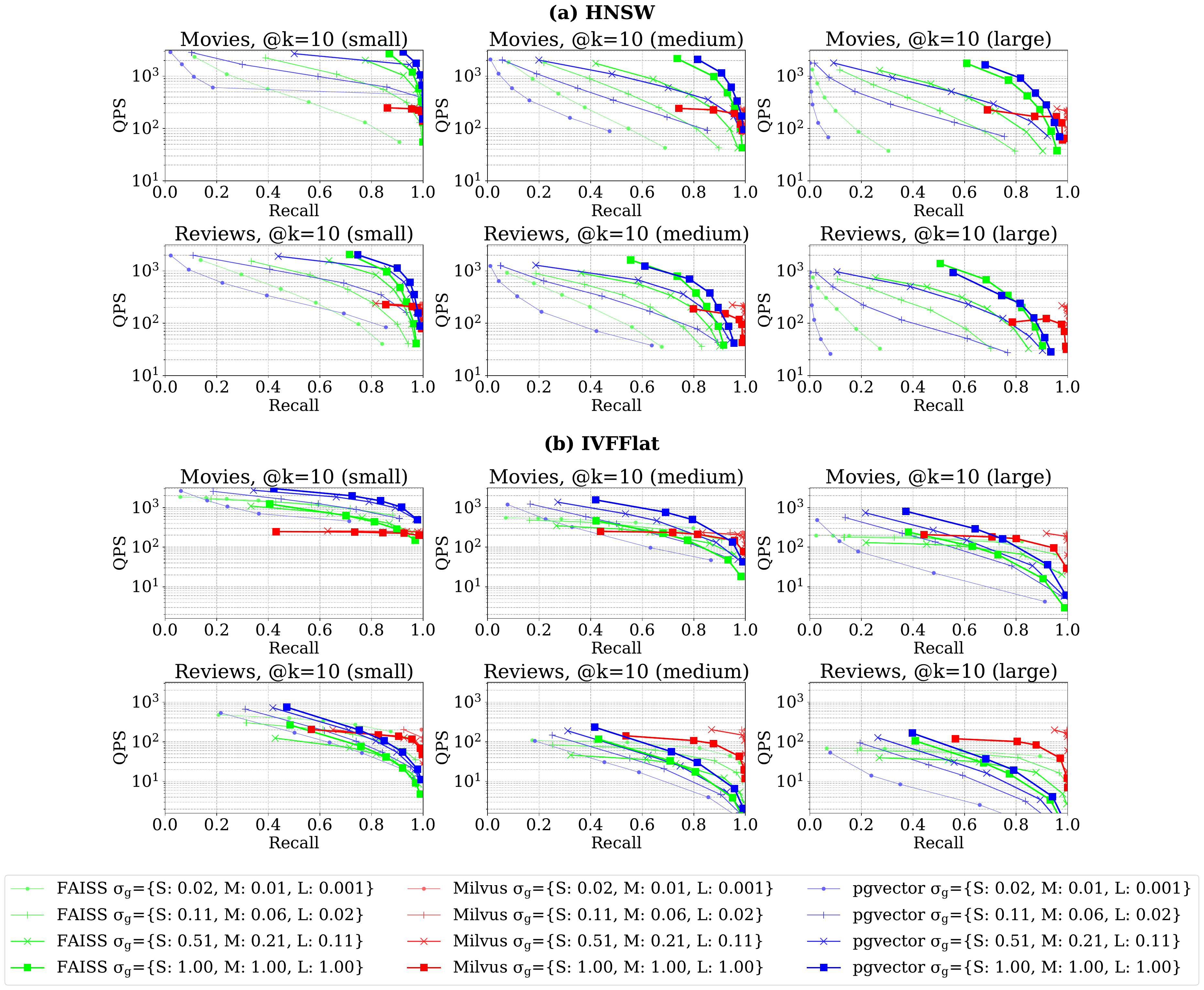}
    \caption{QPS--Recall curves at different selectivity levels.}
    \label{fig:QPS_recall_by_sigma}
\end{figure}

\textbf{Changes to ANN-Benchmarks Implementation.} We extended the ANN-Benchmarks experimental loop to generate YAML configurations dynamically: one ``run group'' per index parameter set during preprocessing. In the query phase, we test filter/$k$/index/search-parameter combinations on the same built index without rebuilds, encoding filters in HDF5 directory hierarchies and aggregating results in a comprehensive CSV (runtimes and recalls per filter/$k$/query/index/search-parameter).

\textbf{Execution Protocol.} Following the ANN-Benchmarks methodology, all queries are executed \textit{single-threaded}, one query at a time. This design isolates the algorithmic and architectural performance characteristics of each system from concurrency effects, enabling controlled comparisons. All evaluated systems operate in \textit{in-memory} mode; indexes and datasets are fully loaded into RAM before query execution begins. We do not evaluate disk-based or streaming workloads. Each query is executed once without repetition within the measurement phase; throughput (QPS) is computed as the inverse of the measured single-query latency. No batch query execution is employed---each query is submitted and completed individually.

\textbf{Ground Truth for Filtered Recall.} For filtered queries, ground truth is computed by first applying the filter predicate to the full dataset, then performing an \textit{exact} $k$-NN search (brute-force distance computation) over the filtered subset. Formally, for a query vector $q$ and filter predicate $\phi$, we define the ground truth as the $k$ vectors from $\{v \in \mathcal{D} \mid \phi(v) = 1\}$ with the smallest distance to $q$.

\textbf{Hardware and Software Configuration.} As in ANN-Benchmarks \cite{aumüller2018_annbenchmarks}, we run all experiments in a Docker container, which includes all necessary dependencies for the evaluated systems. We use a machine running Ubuntu 24.04 LTS equipped with a 56-core Intel Xeon E5-2660 (2.00 GHz) and 256 GB of RAM. Table~\ref{tab:software_versions} lists the exact versions of all evaluated systems. All experiments use the specified versions to ensure reproducibility.

\textbf{Scope.} Our experiments focus exclusively on \textit{scalar inequality filters} (e.g., \texttt{rating >= threshold}). Categorical filters, string pattern matching, and multi-attribute conjunctions are excluded from the experimental workload. Additionally, we evaluate only in-memory execution; disk-based and runtime filtering strategies are out of scope.

\subsection{Results}

\subsubsection{Impact of Filter Selectivity}

As illustrated in Figure~\ref{fig:QPS_recall_by_sigma}, an increase in filter selectivity consistently exerts negative pressure on recall for both HNSW and IVFFlat implementations. This degradation occurs regardless of whether a pre-filtering or post-filtering strategy is employed, though the mechanisms differ. In pre-filtering, low selectivity filters aggressively sparsify the distribution of true nearest neighbors, including the region surrounding the query vector. Conversely, in post-filtering, the effective candidate pool is drastically reduced after the index scan; if the initial search does not retrieve a sufficient number of candidates satisfying the predicate, the final result set is inevitably truncated. \textit{Milvus} is the notable exception, maintaining higher recall at low selectivity due to its hybrid architecture (see Section~\ref{subsubsec:milvus_hybrid}).

Despite yielding slight recall decrease at lower selectivity levels in general, IVFFlat may demonstrate superior QPS performance, as indicated by the elevated QPS-Recall curve in Figure~\ref{fig:QPS_sigma_by_k}. This result is expected, as the cluster-pruning mechanism of IVFFlat reduces the computational overhead of distance calculations.

However, system throughput (QPS) in graph-based indexes remains largely insensitive to variations in selectivity, as the underlying graph structure necessitates consistent traversal efforts regardless of filter presence. This stability indicates that the overhead of filtering is negligible compared to the expensive distance computations required during graph traversal, particularly in large datasets.

\noindent
\begin{minipage}[l]{0.5\textwidth}
\textbf{Predicate evaluation in FAISS}. To satisfy the bitset requirement, we implemented a wrapper around the FAISS index that handles predicate evaluation using NumPy. For every query, we first compute a boolean mask indicating which vectors satisfy the filter, and then convert this mask into a FAISS \texttt{IDSelector}. This process occurs on-the-fly with negligible overhead ($<1\,\text{ms}$ on the large datasets). Figure~\ref{fig:s1_1} summarizes the added latency across representative workloads.
\end{minipage}%
\hfill
\begin{minipage}[r]{0.48\textwidth}
\centering
  \captionsetup{type=figure}
  \includegraphics[width=0.9\textwidth]{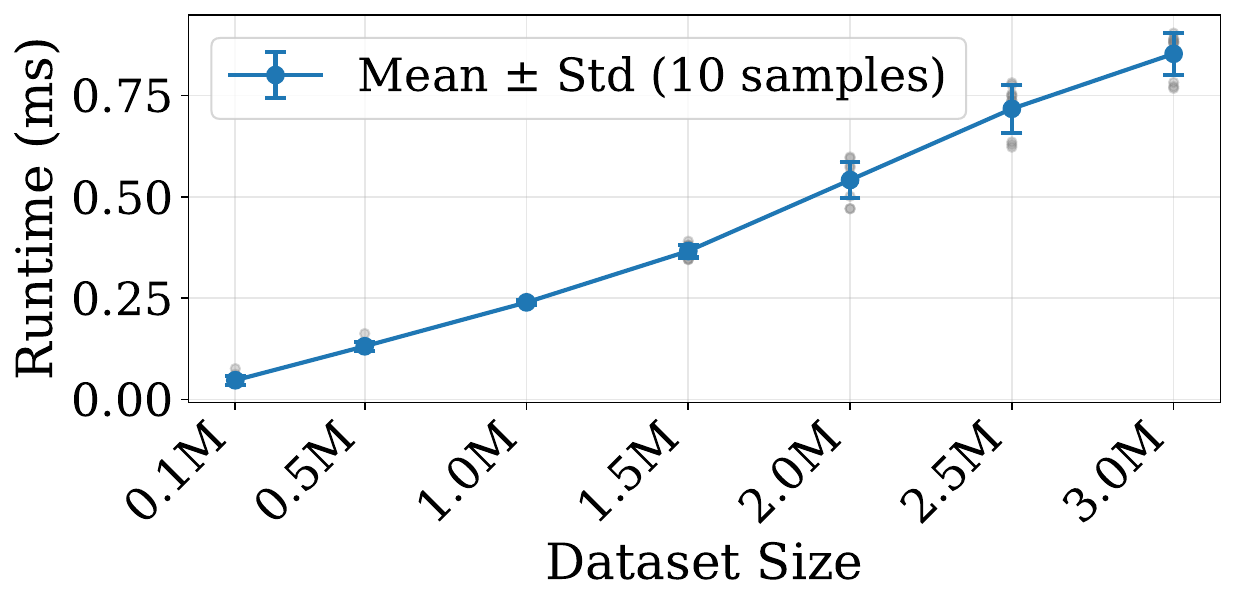}
  \vspace{-1\baselineskip}
  \captionof{figure}{Bitset generation runtime in FAISS.}
  \vspace{0.6\baselineskip}
\label{fig:s1_1}
\end{minipage}

\textbf{Takeaway.} Filter selectivity acts as a universal constraint on recall. However, its impact on throughput varies by architecture: graph-based indices exhibit QPS invariance, confirming that traversal costs mask filtering overheads, whereas partition-based indices (IVFFlat) leverage pruning to achieve superior throughput in low-selectivity regimes. Consequently, performance in filtered ANNS is determined less by metadata evaluation costs and more by the specific mechanics of the chosen vector index—balancing traversal robustness against pruning efficiency.

\subsubsection{Impact of \texorpdfstring{$k$}{k}}

Figure \ref{fig:recall_vs_selectivity_by_k} analyzes the relationship between the number of requested nearest neighbors ($k$) and system recall. Given that search parameters (such as \texttt{efSearch} or \texttt{nprobe}) were held constant, the raw size of the candidate pool generated by the index remains static. As $k$ increases, the system must extract a larger number of valid results from this fixed-size candidate set. This leads to a systematic decrease in recall, as the probability of the fixed set containing $k$ valid neighbors diminishes.

This effect is particularly pronounced in \textit{pgvector}’s HNSW implementation using post-filtering, except when recall jumps to 1.0 at lower selectivities (triggered by the query optimizer switching to a sequential scan). Since filter is applied only to the final set of nearest neighbors bounded by \texttt{efSearch}, there is an increased statistical likelihood that a majority of candidates retrieved by the raw vector search will fail the predicate. Consequently, many true neighbors are never examined because they were not among the top-\texttt{efSearch} candidates in the unfiltered vector space, causing a precipitous drop in recall.

\textbf{Takeaway.} To sustain high recall when increasing the retrieval size $k$, it is imperative to proportionally scale the search parameters (e.g., increasing \texttt{efSearch}). However, this necessitates a trade-off, as larger search parameters incur higher computational costs and increased query latency.

\begin{figure}[htbp]
\centering
  \includegraphics[width=0.9\textwidth]{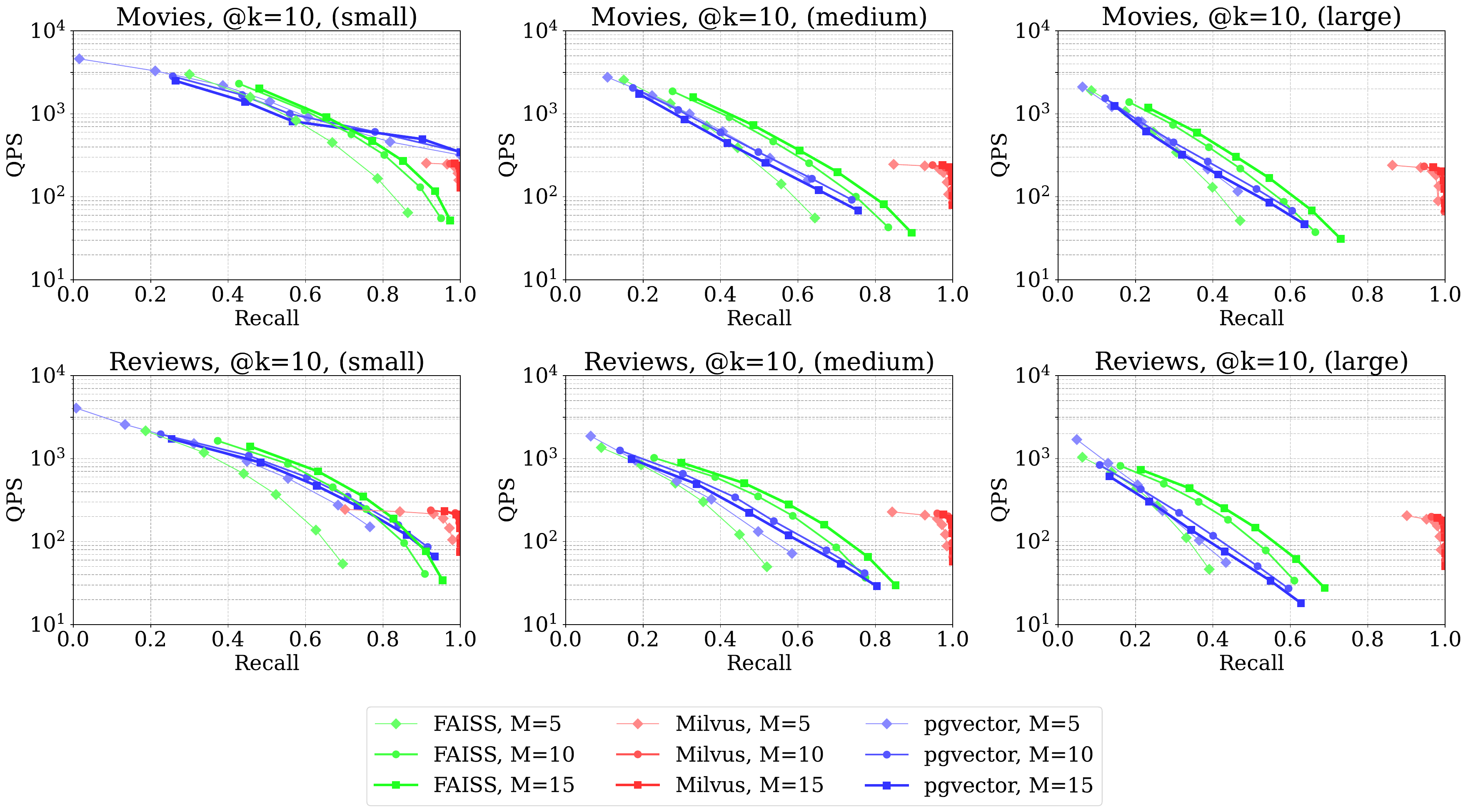}
    \caption{QPS--Recall curves for HNSW at different values of $M$.}
    \label{fig:hnsw_qps_vs_recall_by_m}
\end{figure}

\subsubsection{Filtering Strategies in HNSW}

Post-filtering approach traverses the HNSW graph agnostically, without guidance from the predicate. This results in the admission of numerous candidates that are ultimately discarded, thereby reducing the effective recall achievable within the \texttt{efSearch} budget. Due to that, inferior recall for low-selectivity queries is expected. However, post-filtering implementation in \textit{pgvector} shows comparable and even slightly better performance than pre-filtering implementation of \textit{FAISS} for non-filtered or high-selectivity queries in terms of runtime. We attribute this to \textit{pgvector} avoiding longer graph traversal and early filter evaluation overhead.

Although \textit{Milvus}’s Knowhere engine is built upon \textit{FAISS}, its performance characteristics differ substantially, due to its hybrid architecture (see Section~\ref{subsubsec:milvus_hybrid}).

\textbf{Takeaway.} While pre-filtering is generally expected to provide superior recall, post-filtering may be the optimal choice for filters with high selectivity.

\subsubsection{HNSW-Graph Density}

Increasing the HNSW construction parameter $M$ (and implicitly \texttt{efConstruction}) significantly increases the index building time in \textit{pgvector} and \textit{FAISS}, as shown in Figure \ref{fig:idx_build_time}. Although \textit{Milvus} demonstrates invariance to construction parameters, this also results from its hybrid architecture (Section~\ref{subsubsec:milvus_hybrid}). HNSW inherently incurs higher construction costs with larger $M$ values, as internal vector searches for node insertion become computationally expensive. This increase in graph density yields only negligible gains in recall and has minimal (or even negative) impact on search latency across all systems (Figure~\ref{fig:hnsw_qps_vs_recall_by_m}), unless the graph is excessively sparse (e.g., $M=5$). Therefore, when rapid index construction is prioritized, ``lighter'' graphs with smaller $M$ are preferable.

\begin{minipage}[l]{0.54\textwidth} 
\centering
    \captionsetup{type=figure}
    \includegraphics[width=\textwidth, keepaspectratio]{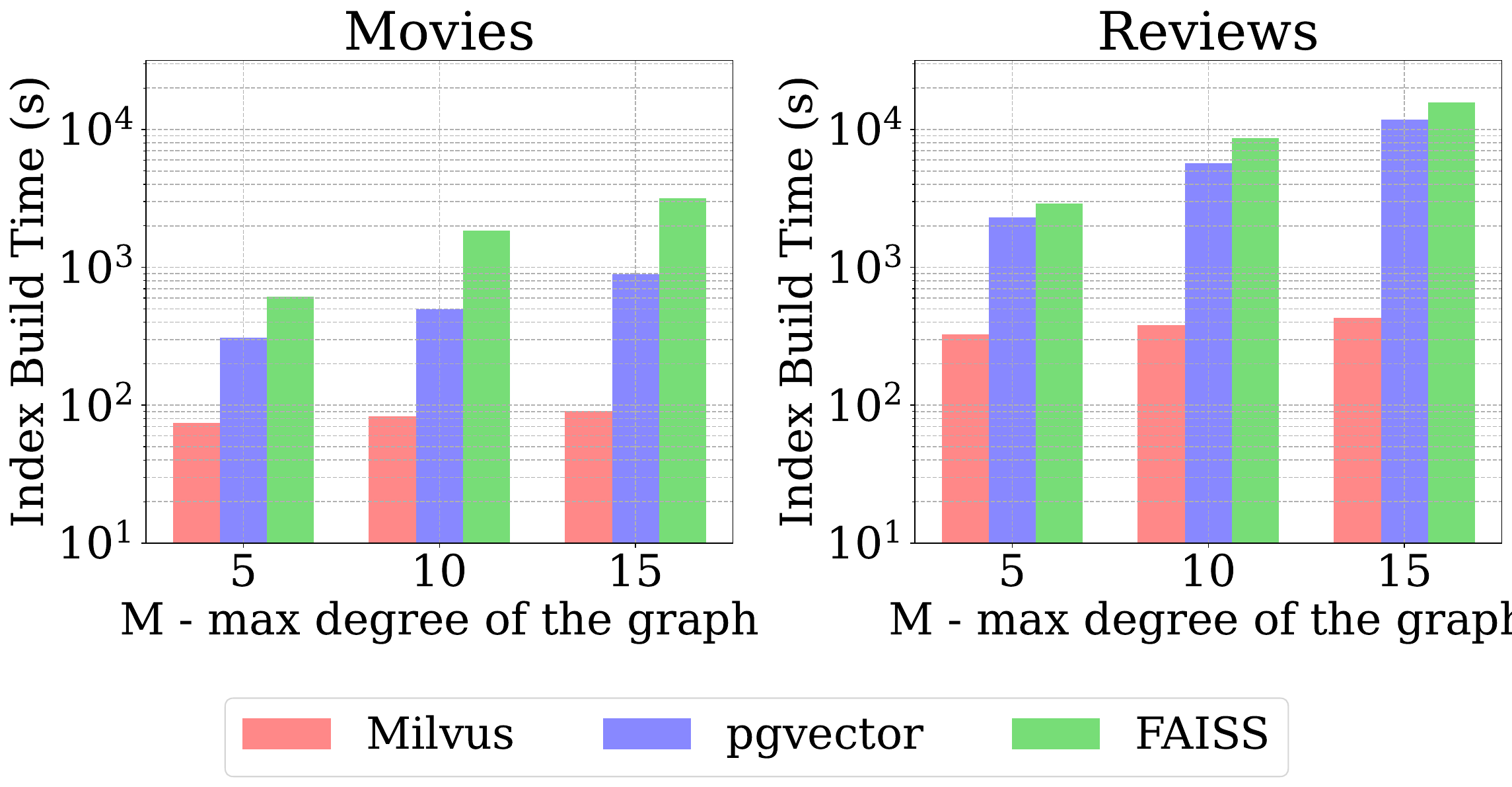}
    \captionof{figure}{Index build times for HNSW.}
    \label{fig:idx_build_time}
\end{minipage}%
\hfill
\begin{minipage}[r]{0.36\textwidth}
\centering
    \captionsetup{type=figure}
    \includegraphics[width=\textwidth, keepaspectratio]{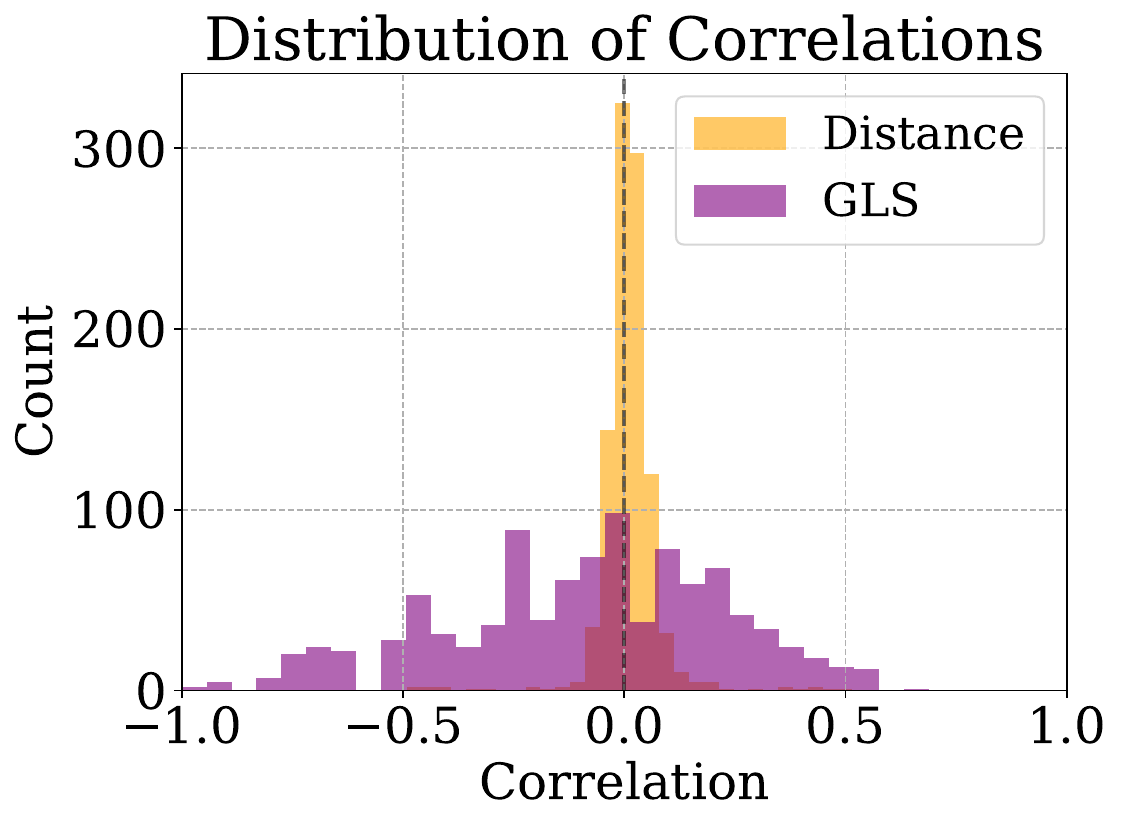}
    \captionof{figure}{Distribution of GLS correlation and distance-based correlation values.}
    \label{fig:gls_vs_acorn_hist}
\end{minipage}

\begin{figure}[htbp]
\centering
  \includegraphics[width=0.89\textwidth, height=\textheight, keepaspectratio]{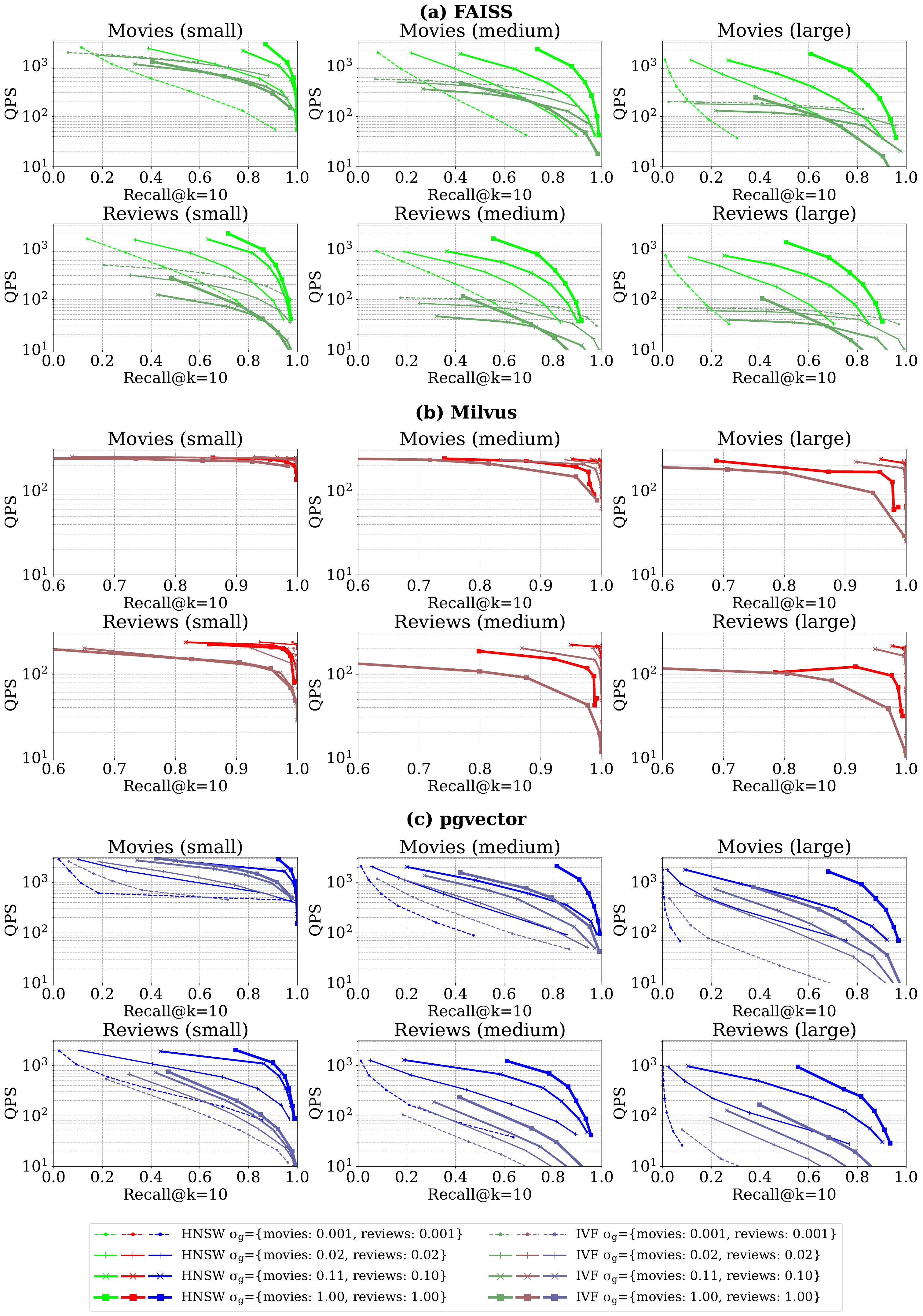}
    \caption{QPS--Recall curves for HNSW and IVFFlat indexes at different selectivity levels.}
    \label{fig:hnsw_vs_ivf}
\end{figure}

\begin{figure}[htbp]
\centering
  \includegraphics[width=0.9\textwidth]{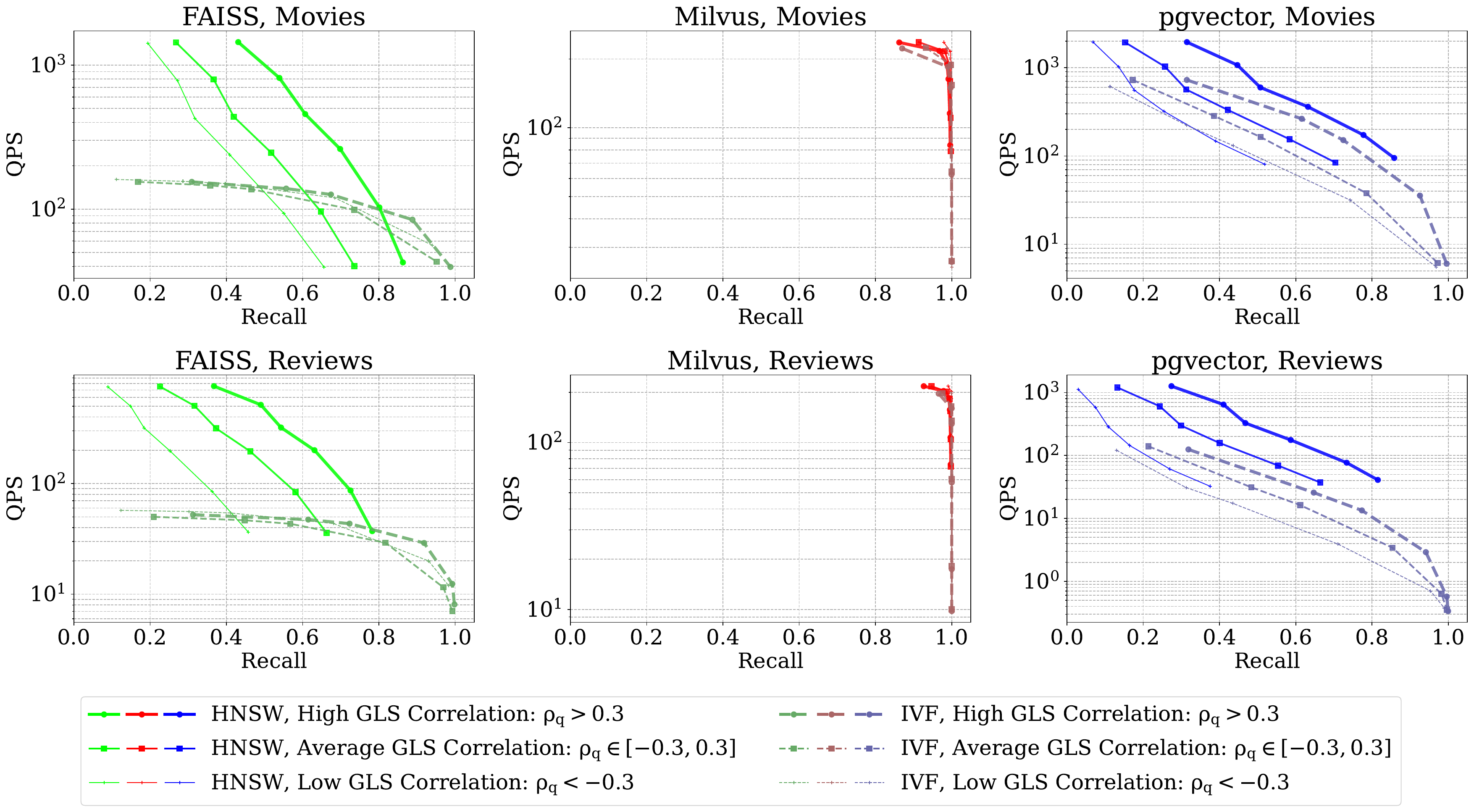}
    \caption{QPS--Recall curves for different levels of GLS correlation.}
    \label{fig:qps_recall_by_GLScorrelation}
\end{figure}

\subsubsection{IVFFlat index}

IVFFlat is traditionally considered inferior to HNSW for general ANNS tasks due to lower recall-throughput efficiency. However, our results indicate specific edge cases—particularly under low selectivity filters—where IVFFlat outperforms HNSW. This is likely due to the cluster-pruning nature of IVFFlat, which may adapt better to sparse regions of the vector space than the graph-traversal mechanics of HNSW.

While HNSW maintains dominance in standard benchmarks, its superiority is not absolute in filtered search. We observed that for each system, specific selectivity thresholds exist where the QPS-Recall curves of HNSW and IVFFlat intersect (Figure \ref{fig:hnsw_vs_ivf}).

\textbf{Takeaway.} HNSW is challenged in low-selectivity scenarios where the sparsity of valid neighbors necessitates traversing distant nodes. This can trigger premature search termination due to lack of updates in the candidate set (see Fig. \ref{fig:o1_1}). In such cases, the partition-based approach of IVFFlat offers a robust alternative, suggesting that: 1) workload-aware index selection can be beneficial for query performance, and 2) hybrid indexing methods may be necessary for optimal performance across the full selectivity spectrum. This may include switching between graph and partition methods based on query predicates, or utilizing adaptable search parameters contingent on estimated selectivity.

\subsubsection{The Impact of Query-Filter Correlation on Performance}

While filter selectivity ($\sigma_g$) serves as a primary driver of performance, our analysis of the \textit{Global-Local Selectivity} (GLS) correlation reveals a more nuanced relationship between metadata and recall. As shown in Figure~\ref{fig:qps_recall_by_GLScorrelation}, retrieval accuracy improves significantly as the GLS correlation shifts from low to high, while query latency (or QPS) remains invariant across all three correlation scenarios. This relationship persists because a high GLS correlation implies that valid neighbors are spatially clustered near the query, making them easier for index structures to locate. Conversely, low or negative GLS correlation indicates that valid neighbors are "pushed" to the periphery of the query's natural neighborhood, significantly challenging both graph-based and partition-based traversal.

Finally, Figure~\ref{fig:gls_vs_acorn_hist} demonstrates the sensitivity of our metric: while distance-based correlations only capture a narrow range of interactions (approximately $[-0.3, 0.3]$), GLS correlation spans nearly the full spectrum from $-1$ to $1$. This confirms that GLS is a more expressive tool for identifying "hard" queries. By categorizing queries into low ($\rho_q < -0.3$), medium ($-0.3 \leq \rho_q \leq 0.3$), and high ($\rho_q > 0.3$) GLS brackets, we show that lower GLS consistently results in lower recall across all tested systems. This does not contradict our previous assertion that selectivity can be studied as an isolated variable for system-level planning; rather, it suggests that while $\sigma_g$ determines the \textit{average} system behavior, GLS correlation explains the \textit{per-query} variance that individual users may experience.

\subsubsection{Milvus: Hybrid Architecture and Adaptive Search}\label{subsubsec:milvus_hybrid}

\begin{figure}
\centering
  \includegraphics[width=0.7\textwidth]{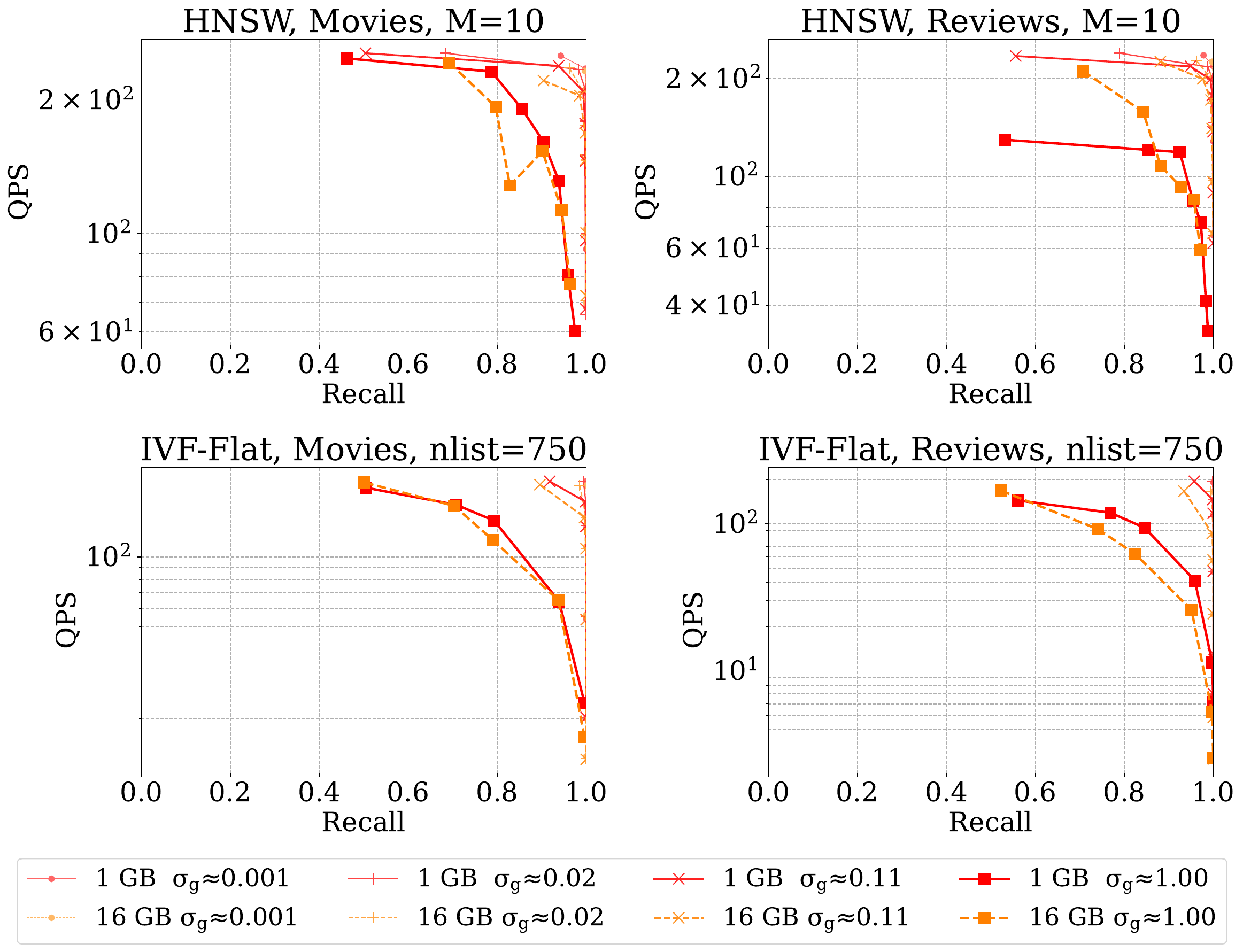}
  \caption{QPS--Recall curves for different segment sizes in Milvus.}
  \label{fig:ablation_milvus}
\end{figure}

The \textit{Milvus} vector database operates on a specialized engine called \textit{Knowhere}, which diverges from standard FAISS wrappers to ensure stability during filtered search. \textit{Knowhere} employs FAISS for graph construction but executes all search operations through a heavily modified fork of \textit{hnswlib} \cite{hnswlib}. Contrary to our initial hypothesis, the system's high recall is not primarily driven by data segmentation or "Scatter-Gather" mechanics. Our ablation studies demonstrate that \textit{Milvus} maintains near perfect recall regardless of whether the data is partitioned into many 1GB segments or stored in a single 16GB monolith (Figure \ref{fig:ablation_milvus}). Instead, this robustness stems from a layered algorithmic strategy within the search engine itself.

\textbf{Dual-Pool Graph Traversal.} The most critical differentiator is Knowhere's handling of the candidate queue during graph traversal. Standard FAISS HNSW implementations use a single priority queue where filtered (invalid) nodes compete for slots with valid nodes. At high filter rates (e.g., $90\%$ filtered out), the search beam becomes saturated with invalid navigation nodes, causing recall to plummet. Knowhere addresses this by implementing a \textit{Dual-Pool} strategy (\texttt{NeighborSetDoublePopList}). It maintains two separate priority queues during traversal: one for valid results and another for invalid nodes used strictly for navigation. This ensures that the search budget ($ef$) is not cannibalized by filtered vectors, allowing \textit{Milvus} to maintain near-perfect recall even at moderate selectivities (e.g., $10\%--5\%$) where standard HNSW implementations degrade.

\textbf{Adaptive Fallback Mechanisms.} To handle highly restrictive filters, \textit{Milvus} employs a deterministic fallback strategy rather than a cost-based optimizer. The engine checks filter selectivity prior to search; if the ratio of filtered vectors exceeds a specific threshold (set to $93\%$ in the codebase), it bypasses the HNSW index entirely and performs a brute-force scan (Pre-filtering + kNNS). Given the reduced search space, this operation remains computationally inexpensive (1--5ms for typical segments) while guaranteeing $100\%$ recall. Furthermore, a "safety net" fallback is triggered post-search if the graph traversal fails to return the requested $k$ neighbors, ensuring consistency even in sparse vector regions.

This adaptive behavior extends to \textit{IVFFlat} indexes, explaining the anomaly where \textit{Milvus} achieves 1.0 recall even with $nprobe=1$ under restrictive filters. While standard inverted indices fail in this regime because the few matching vectors are scattered across unvisited clusters, Knowhere detects the high selectivity and switches to a linear scan of the valid vectors, effectively treating the query as a small-data problem. Consequently, \textit{Milvus} exhibits "Exact Search" behavior for highly filtered queries regardless of the underlying index type.

\textbf{Construction Efficiency.} The robustness of index construction is driven by a batch-parallel build pipeline using a persistent thread pool (Folly), rather than simple segment parallelization. Knowhere also performs a post-build graph repair step to reconnect vectors isolated during concurrent insertion, which explains the higher base recall compared to vanilla FAISS.

\textbf{Attribute Index.} \textit{Milvus} performance remains less sensitive to the presence of an attribute index (Figure \ref{fig:qps_vs_recall_attidx_milvus}). This contrasts sharply with \textit{pgvector}, which may change its query execution plan depending on configurations. Our analysis confirms that \textit{Milvus} does not employ a cost-based Query Optimizer in the traditional relational sense, instead relying on this static pre-filtering and per-segment fallback mechanism.

\textbf{Takeaway.} Milvus prioritizes recall stability through a sophisticated "hybrid" engine design. Its performance profile is a byproduct of its hybridized algorithmic specialization: the \textit{Dual-Pool} traversal prevents search starvation under moderate filters, while adaptive brute-force fallbacks guarantee accuracy under strict filters for both HNSW and IVFFlat. This design allows the system to behave like an exact search engine, delivering perfect recall for queries with high selectivity and hard-to-find neighbors.

\subsubsection{PG-Vector: Query Optimization for Filtered ANNS}

\textbf{Falling back to kNN.} The PostgreSQL Query Optimizer is designed to select the execution plan with the lowest estimated total cost. To integrate vector similarity search, \textit{pgvector} introduces specific cost functions for each supported vector index type. These calculations rely heavily on cardinality estimates—standard in relational DBMS optimizers—to predict the number of tuples involved in the ANNS operation. The overall cost function is sensitive to the parameter $k$ and the estimated cardinality of the dataset following the application of the selection predicate $S_{\sigma}$. We observed two distinct query plans:

\begin{itemize}
    \item \textit{ANNS + Post-filtering:} In this plan, the vector search on index is executed first. For HNSW, this involves populating a queue of size $ef_{search}$, while for IVFFlat, it involves probing $n_p$ clusters. Selection predicates are subsequently applied to the tuples returned by the index scan. Below is an illustrative plan for a query with $r = 9.5$ and \textbf{$k=20$} using the HNSW index:
    
\begin{small}
\begin{verbatim}
Limit
  ->  Index Scan using vector_idx on dataset
        Order By: (vector <=> $0)
        Filter: (attribute > 8.5)
\end{verbatim}
\end{small}

    \item \textit{Pre-Filtering + kNNS:} Here, the selection predicates are applied first via a sequential scan, followed by an exact k-NN search performed by calculating distances between the query vector and the filtered tuples. While this fallback to brute-force search guarantees theoretically perfect recall (Exact kNN), it can be computationally suboptimal for larger subsets. Below is the execution plan for the query with $r = 9.5$ and \textbf{$k=21$}:

\begin{small}
\begin{verbatim}
Limit 
  ->  Sort
        Sort Key: ((dataset.vector <=> $0))"
        Sort Method: top-N heapsort  Memory: XYZ kB"
        ->  Seq Scan on dataset 
              Filter: (attribute > 8.5)"
\end{verbatim}
\end{small}
\end{itemize}

We subsequently replicated the experiment with an active B-Tree index on the metadata attribute used for filtering. In this configuration, the query optimizer is presented with an additional execution path:
\begin{itemize}
    \item \textbf{Attribute Index Scan + kNNS}: This plan leverages the index on \verb|averagerating| to filter the dataset efficiently, after which it identifies the $k$-nearest neighbors by strictly calculating distances to the query vector. Below is the execution plan for the query with $r = 9.5$ and \textbf{$k = 10$}:
\end{itemize}
        
    \begin{small}
    \begin{verbatim}
Limit
  ->  Sort
        Sort Key: ((dataset.vector <=> $0))
        Sort Method: top-N heapsort
        ->  Bitmap Heap Scan on dataset
              Recheck Cond: (attribute > 8.5)
              ->  Bitmap Index Scan on idx_attribute
                    Index Cond: (attribute > 8.5::double precision)
    \end{verbatim}
    \end{small}

As illustrated in Figure \ref{fig:qps_vs_recall_attidx_pgvector}, the \textit{Attribute Index Scan + kNNS} plan consistently yields perfect recall. However, the automatic plan selection in \textit{pgvector} appears suboptimal; for many low-selectivity queries, the query optimizer overlooks sequential scan based plans in favor of vector search on indexes. Our data indicates that there are cases, when forcing the \textit{Attribute Index Scan + kNNS} (or \textit{Pre-Filtering + kNNS}) plan would achieve perfect recall at latencies comparable to the selected approximate plans.

\textbf{Takeaway}: Indexing metadata in \textit{pgvector} significantly alters query execution planning, yet exposes limitations in the optimizer's cost model. Specifically, the optimizer often prioritizes approximate vector index searches even in scenarios where filtering followed by exact distance computation would offer guaranteed recall at a similar or lower computational cost. This suggests that the cost weighting between high-dimensional vector operations and scalar index scans requires further tuning to avoid recall-suboptimal plan selections.

\subsection{Summary}

Our extensive empirical evaluation of several open-source systems, including \textit{FAISS}, \textit{Milvus}, and \textit{pgvector}, yields several critical insights into the interplay between filter selectivity, index topology, and system architecture:

\textbf{The Selectivity-Efficiency Trade-off:} We observed divergent behavior across index types. While graph-based indexes exhibit degrading QPS-Recall curves at low selectivities regardless of filtering strategy, partition-based indexes (IVFFlat) demonstrate significant efficiency gains in some cases. As such, in FAISS, the pre-filtering bitset allows the search to bypass expensive distance computations for invalid vectors within probed clusters. Although absolute recall at a fixed $n_{\text{probe}}$ may decrease due to a constrained candidate pool, the reduced computational overhead drastically increases throughput (QPS). This results in a superior QPS-Recall trade-off curve of IVFFlat compared to higher selectivity scenarios.

\textbf{Index Inversion:} We demonstrated that the traditional dominance of HNSW is challenged in low-selectivity scenarios. In such cases, partition-based indexes (IVFFlat) can offer greater robustness due to their efficient cluster-pruning mechanics.

\textbf{Algorithmic Robustness and the Latency Floor:} Our analysis of \textit{Milvus} reveals that recall stability is achieved through algorithmic specialization across \textit{both} graph and partition-based indexes. The system employs adaptive brute-force fallbacks and "Dual-Pool" traversal to achieve high recall for both HNSW and IVFFlat. However, this robustness effectively enforces a "latency floor," sacrificing peak throughput to maintain exact-search accuracy under strict filters.

\textbf{Optimizer Fragility:} We exposed some limitations in \textit{pgvector}'s cost-based optimizer, which frequently fails to switch to sequential scans or attribute-index scans when they would provide perfect recall at competitive latencies. This highlights the need for better cardinality and cost estimation models in relational vector extensions.

\textbf{Query-Filter correlation}: Our findings suggest that while metadata and vector distributions are weakly correlated on average (mean GLS $\approx 0$), individual query performance is highly sensitive to local correlation shifts. Using our GLS metric, we demonstrate that higher GLS correlation serves as a predictor for increased recall, as it indicates a spatial clustering of valid neighbors. However, the high variance and near-zero mean across the dataset suggest that while correlation explains per-query recall fluctuations, it remains too volatile for global cost estimation. Consequently, query optimizers can reliably use global selectivity for plan selection, but must account for the fact that low-correlation "hard" queries may require more aggressive search parameters to maintain target recall levels.

\subsection{Practical Guidelines for Filtered Vector Search}

Based on our experimental findings, we distill the following actionable recommendations for practitioners deploying filtered vector search systems:

\begin{enumerate}[leftmargin=*]
    \item \textbf{Select index type based on expected filter selectivity.} Do not assume HNSW universally dominates IVFFlat. For workloads where expected selectivity $\sigma_g \lesssim 5\%$, consider IVFFlat; for $\sigma_g \gtrsim 20\%$ or predominantly unfiltered queries, HNSW remains preferable. For mixed workloads, maintain both index types and select per query when possible.
    
    \item \textbf{Plan for recall degradation under filtering.} Filtered ANNS inherently degrades recall as selectivity decreases---regardless of system or index. Treat recall loss as a budgeted resource. For critical queries, expose search-parameter overrides or allow fallback to exact search when candidate exhaustion is detected.
    
    \item \textbf{Scale search parameters with $k$.} Since the candidate pool size is fixed by search parameters (\texttt{efSearch}, \texttt{nprobe}), increasing $k$ without scaling these parameters systematically reduces recall. Scale \texttt{efSearch} $\propto k$ (or \texttt{nprobe} $\propto k$) when recall matters.
    
    \item \textbf{Expect a throughput penalty for robustness.} Engines that guarantee high recall under filtering (e.g., \textit{Milvus}) prevent the fail-fast'' behavior seen in standard libraries. While this ensures accuracy, it precludes the ultra-high QPS observed in systems that allow recall to degrade (e.g., \textit{FAISS}, \textit{pgvector}). Practitioners transitioning to robust engines should budget for this latency floor, as the system will refuse to return fast, low-quality results.
    
    \item \textbf{Avoid over-tuning graph density.} Increasing HNSW construction parameters ($M$, \texttt{efConstruction}) yields diminishing returns under filtering, with marginal recall gains but significant build-time increases. Prefer lighter graphs and allocate tuning budget to search-time parameters.
    
    \item \textbf{Index metadata attributes aggressively (pgvector).} In relational vector extensions, metadata indexes unlock execution plans that the optimizer can exploit for filtered queries. Without them, the system often commits to recall-degrading post-filtering strategies.
    
    \item \textbf{Do not blindly trust query optimizers.} Verify query plans manually (\texttt{EXPLAIN ANALYZE}) for critical workloads. The optimizer may favor approximate index scans even when exact scans would yield perfect recall at comparable latency. For very selective filters, consider forcing attribute-index or sequential scan plans.
    
    \item \textbf{Use global selectivity for plan selection, but monitor correlation.} Since metadata attributes tend to be weakly correlated with the embedding space on average (mean GLS $\approx 0$), system throughput is dominated by global selectivity ($\sigma_g$). Base primary plan selection on $\sigma_g$ and $k$: low $\sigma_g$ favors exact or pre-filtered search, while high $\sigma_g$ makes ANN-first plans safe. However, be aware that queries with low GLS correlation (where valid neighbors are "pushed" away from the query) will suffer from lower recall. In recall-critical applications, consider using GLS as a diagnostic tool to identify these "hard" query regions and dynamically increase search depth (e.g., $ef_{\text{search}}$).
\end{enumerate}

\section{CONCLUSIONS AND FUTURE WORK}\label{sec:conclusions}

In this work, we addressed the widening gap between algorithmic theory and system-level implementation in FANNS. First, we systematized the ambiguous landscape of FANNS by formalizing a taxonomy of Pre-, Post-, and Runtime-filtering for both graph-based and partition-based indexes. Second, we established a rigorous evaluation framework for modern Vector Databases handling predicate-based queries. To support this, we introduced MoReVec, a new dataset comprising two relations: \textit{Movies} and \textit{Reviews}. These relations are connected via a foreign key and feature transformer-based embeddings alongside structured metadata attributes. Finally, we proposed the Global-Local Selectivity (GLS) metric to quantify the independence between metadata attributes and vector embeddings—a critical distinction for isolating the impact of selectivity from semantic clustering.

\textbf{Limitations.} We acknowledge several factors that may affect the generalizability of our findings:

\textit{Internal Validity.} Our conclusions are sensitive to the specific hyperparameter ranges explored (\texttt{efSearch}, \texttt{nprobe}, $M$). Different parameter sweeps may yield different crossover points between index types. All experiments are conducted in-memory with datasets fully loaded into RAM; results may differ under memory pressure or disk-based execution. Our evaluation inherits the ANN-Benchmarks single-threaded execution model, which isolates algorithmic behavior but does not capture production concurrency effects.

\textit{External Validity.} Results may not generalize to: (1) other embedding models (e.g., CLIP, E5, OpenAI embeddings) that produce different vector distributions; (2) alternative distance metrics (inner product, $L_2$) on non-normalized vectors; (3) workloads with highly correlated metadata, where filter attributes are semantically aligned with the embedding space; and (4) distributed or multi-node deployments, where network latency and data partitioning introduce additional variables. Our conclusions apply primarily to generic, schema-agnostic filtering strategies within single-node Vector Databases---not to fusion methods (e.g., ACORN, HQANN) or workload-aware indexes.

\textbf{Future Work.} The integration of vector search into relational systems remains a frontier for database research. While current systems like pgvector support hybrid queries syntactically, their optimizers often lack the sophistication to maintain performance under varied workloads. We identify the development of a Selectivity-Aware Query Optimizer for Filtered-ANNS as a primary direction. Such an optimizer must go beyond simple heuristics to perform:

\begin{itemize}
    \item \textit{Optimal Index \& Parameter Selection}: Dynamically choosing between graph-based (e.g., HNSW) or partition-based (e.g., IVF) indices, and tuning search parameters (e.g., $ef_{search}$, $n_{probe}$) at runtime based on specific filter constraints and budget.
    
    \item \textit{Correlation-Aware Estimates}: Investigating estimated GLS correlations to predict the "reachability" of filter-passing neighbors. This allows the engine to account for cases where attribute distributions are non-uniformly clustered in the vector space, impacting the search effort required to identify valid candidates.
    
    \item \textit{Complex Relational Interops}: Extending optimization logic to JOIN + Vector Search scenarios, when immediate attribute filtering is infeasible. In such cases the optimizer must evaluate not only the cost-efficiency of competing execution plans, but also estimated Recall of competing execution plans: (1) Join-First, where the relational join is executed to narrow the candidate pool prior to vector search; or (2) ANNS-First, which performs a vector search to retrieve a speculatively oversized candidate set, followed by a late-stage join to resolve remaining relational constraints.
\end{itemize}

Additionally, we intend to investigate hybrid indexing strategies that integrate multiple algorithmic approaches (e.g., graph-structured partitions) into a single, unified design. In this framework, real-time selectivity estimates do not trigger a switch between disparate indices but rather guide internal traversal logic and parameters. Such an index is intended to remain robust under restrictive filter constraints and the distribution shifts typical of evolving relational datasets.

\section*{Acknowledgments}
This work was supported by NSF award number 2008815.

\small{
 \bibliographystyle{abbrv}
 \bibliography{biblio}

@misc{annbenchmark_extension,
  key = {ANN Benchmarks Extension},
  title = {{ANN Benchmarks Extension for Filtered Search}},
  howpublished = "\url{https://github.com/aabylay/ANN-benchmark-HQ}",
  note = {Accessed: 2026-01-20}
}

@misc{faiss,
  key = {FAISS},
  title = {{FAISS: A Library for Efficient Similarity Search}},
  howpublished = "\url{https://github.com/facebookresearch/faiss}",
  note = {Accessed: 2026-01-20}
}

@misc{pgvector4aws,
  key = {pgvector AWS},
  title = {{Supercharging vector search performance and relevance with pgvector 0.8.0 on Amazon Aurora PostgreSQL}},
  howpublished = "{\url{https://aws.amazon.com/blogs/database/supercharging-vector-search-performance-and-relevance-with-pgvector-0-8-0-on-amazon-aurora-postgresql/}}",
  note = {Accessed: 2026-01-20}
}

@misc{pgvector,
  key = {pgvector},
  title = {{pgvector: Open-source vector similarity search for Postgres}},
  howpublished = "\url{https://github.com/pgvector/pgvector}",
  note = {Accessed: 2026-01-20}
}

@misc{annbenchmarks,
  key = {ANN-Benchmarks},
  title = {{ANN-Benchmarks: A Benchmarking Tool for Approximate Nearest Neighbor Algorithms}},
  howpublished = "\url{https://ann-benchmarks.com/}",
  note = {Accessed: 2026-01-20}
}

@misc{hnswlib,
  key = {hnswlib},
  title = {{Header-only C++/python library for fast approximate nearest neighbors}},
  howpublished = "\url{https://github.com/nmslib/hnswlib}",
  note = {Accessed: 2026-02-09}
}

@misc{vecdbbenchmark,
  key = {VecDB Benchmark},
  title = {{Vector Database Benchmarks}},
  howpublished = "\url{https://zilliz.com/vdbbench-leaderboard}",
  note = {Accessed: 2026-01-20}
}

@misc{milvus,
  key = {Milvus},
  title = {{Milvus: A Cloud-Native Vector Database}},
  howpublished = "\url{https://milvus.io/}",
  note = {Accessed: 2026-01-20}
}

@misc{pinecone,
  key = {Pinecone},
  title = {{Pinecone: Vector Database for Machine Learning}},
  howpublished = "\url{https://www.pinecone.io/}",
  note = {Accessed: 2026-01-20}
}

@misc{weaviate,
  key = {Weaviate},
  title = {{Weaviate: Open-source Vector Database}},
  howpublished = "\url{https://github.com/weaviate}",
  note = {Accessed: 2026-01-20}
}

@misc{vespa,
  key = {Vespa},
  title = {{Vespa: The Open Big Data Serving Engine}},
  howpublished = "\url{https://vespa.ai/}",
  note = {Accessed: 2026-01-20}
}

@misc{qdrant,
  key = {Qdrant},
  title = {{Qdrant: Vector Similarity Search Engine and Vector Database}},
  howpublished = "\url{https://github.com/qdrant/qdrant}",
  note = {Accessed: 2026-01-20}
}

@misc{elasticsearch,
  key = {Elasticsearch},
  title = {{Elasticsearch: Free and Open Source, Distributed, RESTful Search Engine}},
  howpublished = "\url{https://github.com/elastic/elasticsearch}",
  note = {Accessed: 2026-01-20}
}

@misc{annoy,
  key = {Annoy},
  title = {{Annoy: Approximate Nearest Neighbors Oh Yeah}},
  howpublished = "\url{https://github.com/spotify/annoy}",
  note = {Accessed: 2026-01-20}
}

@misc{mteb_leaderboard,
  key = {MTEB},
  title = {{Massive Text Embedding Benchmark Leaderboard}},
  howpublished = "\url{https://huggingface.co/spaces/mteb/leaderboard}",
  note = {Accessed: 2026-01-20}
}

@misc{anndatasets,
  key = {ANN Datasets},
  title = {{Evaluation of Approximate Nearest Neighbors: Large Datasets}},
  howpublished = "\url{https://corpus-texmex.irisa.fr/}",
  note = {Accessed: 2026-01-20}
}

@article{faiss2025_vecdb,
  author={Douze, Matthijs and Guzhva, Alexandr and Deng, Chengqi and Johnson, Jeff and Szilvasy, Gergely and Mazaré, Pierre-Emmanuel and Lomeli, Maria and Hosseini, Lucas and Jégou, Hervé},
  journal={IEEE Transactions on Big Data}, 
  title={THE FAISS LIBRARY}, 
  year={2025},
  volume={},
  number={},
  pages={1-17},
  doi={10.1109/TBDATA.2025.3618474}
}

@article{iff_fann_bench2025_swiss,
      title={Benchmarking Filtered Approximate Nearest Neighbor Search Algorithms on Transformer-based Embedding Vectors}, 
      author={Patrick Iff and Paul Bruegger and Marcin Chrapek and Maciej Besta and Torsten Hoefler},
      year={2025},
      journal = {CoRR},
      volume = {arXiv:2507.21989},
}

@article{fann_bench2025_china_fudan,
      title={Filtered Approximate Nearest Neighbor Search: A Unified Benchmark and Systematic Experimental Study [Experiment, Analysis \& Benchmark]}, 
      author={Jiayang Shi and Yuzheng Cai and Weiguo Zheng},
      year={2025},
      eprint={2509.07789},
      archivePrefix={arXiv},
      primaryClass={cs.DB},
      url={https://arxiv.org/abs/2509.07789}, 
      journal = {CoRR},
      volume = {arXiv:2509.07789},
}

@article{fann_bench2025_li,
      title={Attribute Filtering in Approximate Nearest Neighbor Search: An In-depth Experimental Study}, 
      author={Mocheng Li and Xiao Yan and Baotong Lu and Yue Zhang and James Cheng and Chenhao Ma},
      year={2025},
      eprint={2508.16263},
      archivePrefix={arXiv},
      primaryClass={cs.DB},
      url={https://arxiv.org/abs/2508.16263}, 
      journal = {CoRR},
      volume = {arXiv:2508.16263},
}

@article{lin2025_survey_fanns,
      title={Survey of Filtered Approximate Nearest Neighbor Search over the Vector-Scalar Hybrid Data}, 
      author={Yanjun Lin and Kai Zhang and Zhenying He and Yinan Jing and X. Sean Wang},
      year={2025},
      eprint={2505.06501},
      archivePrefix={arXiv},
      primaryClass={cs.DB},
      url={https://arxiv.org/abs/2505.06501}, 
      journal = {CoRR},
      volume = {arXiv:2505.06501},
}

@article{xia2026_fannsQO,
    title={Filtered Approximate Nearest Neighbor Search Cost Estimation},
    author={Wenxuan Xia and Mingyu Yang and Wentao Li and Wei Wang},
    year={2026},
    eprint={2602.06721},
    archivePrefix={arXiv},
    primaryClass={cs.DB},
    url={https://arxiv.org/pdf/2602.06721},
    journal = {CoRR},
    volume = {arXiv:2602.06721},
}

@inproceedings{yunan2024_rdbms_vs_vecdb,
  author={Zhang, Yunan and Liu, Shige and Wang, Jianguo},
  booktitle={2024 IEEE 40th International Conference on Data Engineering (ICDE)}, 
  title={Are There Fundamental Limitations in Supporting Vector Data Management in Relational Databases? A Case Study of PostgreSQL}, 
  year={2024},
  volume={},
  number={},
  pages={3640-3653},
  doi={10.1109/ICDE60146.2024.00280}}

@article{li2019_ann_tkde,
  author = {Li, Wen and Zhang, Yong and Wang, Hongya and Lu, Kejing and Kudo, Mineichi},
  title = {Approximate Nearest Neighbor Search on High Dimensional Data: Experiments, Analyses, and Improvement},
  journal = {IEEE Transactions on Knowledge and Data Engineering},
  volume = {32},
  number = {8},
  pages = {1475--1488},
  year = {2020},
  doi = {10.1109/TKDE.2019.2909204}
}

@article{simhadri2024_results_bigann,
      title={Results of the Big ANN: NeurIPS'23 competition}, 
      author={Harsha Vardhan Simhadri and Martin Aumüller and Amir Ingber and Matthijs Douze and George Williams and Magdalen Dobson Manohar and Dmitry Baranchuk and Edo Liberty and Frank Liu and Ben Landrum and Mazin Karjikar and Laxman Dhulipala and Meng Chen and Yue Chen and Rui Ma and Kai Zhang and Yuzheng Cai and Jiayang Shi and Yizhuo Chen and Weiguo Zheng and Zihao Wan and Jie Yin and Ben Huang},
      year={2024},
      eprint={2409.17424},
      archivePrefix={arXiv},
      primaryClass={cs.IR},
      url={https://arxiv.org/abs/2409.17424}, 
      journal = {CoRR},
      volume = {arXiv:2409.17424},
}

@article{jin2026_curator_filter,
      title={Curator: Efficient Vector Search with Low-Selectivity Filters}, 
      author={Yicheng Jin and Yongji Wu and Wenjun Hu and Bruce M. Maggs and Jun Yang and Xiao Zhang and Danyang Zhuo},
      year={2026},
      eprint={2601.01291},
      archivePrefix={arXiv},
      primaryClass={cs.DB},
      url={https://arxiv.org/abs/2601.01291}, 
      journal = {CoRR},
      volume = {arXiv:2601.01291},
}

@article{malkov2018_hnsw,
      title={Efficient and robust approximate nearest neighbor search using Hierarchical Navigable Small World graphs}, 
      author={Yu. A. Malkov and D. A. Yashunin},
      year={2018},
      eprint={1603.09320},
      archivePrefix={arXiv},
      primaryClass={cs.DS},
      url={https://arxiv.org/abs/1603.09320}, 
      journal = {CoRR},
      volume = {arXiv:1603.09320},
}

@article{aumüller2018_annbenchmarks,
      title={ANN-Benchmarks: A Benchmarking Tool for Approximate Nearest Neighbor Algorithms}, 
      author={Martin Aumüller and Erik Bernhardsson and Alexander Faithfull},
      year={2018},
      eprint={1807.05614},
      archivePrefix={arXiv},
      primaryClass={cs.IR},
      url={https://arxiv.org/abs/1807.05614}, 
      journal = {CoRR},
      volume = {arXiv:1807.05614},
}

@inproceedings{subramanya_NEURIPS2019_diskann,
 author = {Jayaram Subramanya, Suhas and Devvrit, Fnu and Simhadri, Harsha Vardhan and Krishnawamy, Ravishankar and Kadekodi, Rohan},
 booktitle = {Advances in Neural Information Processing Systems},
 editor = {H. Wallach and H. Larochelle and A. Beygelzimer and F. d\textquotesingle Alch\'{e}-Buc and E. Fox and R. Garnett},
 pages = {},
 publisher = {Curran Associates, Inc.},
 title = {DiskANN: Fast Accurate Billion-point Nearest Neighbor Search on a Single Node},
 url = {https://proceedings.neurips.cc/paper_files/paper/2019/file/09853c7fb1d3f8ee67a61b6bf4a7f8e6-Paper.pdf},
 volume = {32},
 year = {2019}
}

@inproceedings{filtered-diskann,
author = {Gollapudi, Siddharth and Karia, Neel and Sivashankar, Varun and Krishnaswamy, Ravishankar and Begwani, Nikit and Raz, Swapnil and Lin, Yiyong and Zhang, Yin and Mahapatro, Neelam and Srinivasan, Premkumar and Singh, Amit and Simhadri, Harsha Vardhan},
title = {Filtered-DiskANN: Graph Algorithms for Approximate Nearest Neighbor Search with Filters},
year = {2023},
isbn = {9781450394161},
publisher = {Association for Computing Machinery},
address = {New York, NY, USA},
url = {https://doi.org/10.1145/3543507.3583552},
doi = {10.1145/3543507.3583552},
booktitle = {Proceedings of the ACM Web Conference 2023},
pages = {3406–3416},
numpages = {11},
location = {Austin, TX, USA},
series = {WWW '23}
}

@ARTICLE{jegou2011_pq,
  author={Jégou, Herve and Douze, Matthijs and Schmid, Cordelia},
  journal={IEEE Transactions on Pattern Analysis and Machine Intelligence}, 
  title={Product Quantization for Nearest Neighbor Search}, 
  year={2011},
  volume={33},
  number={1},
  pages={117-128},
  doi={10.1109/TPAMI.2010.57}}

@article{bentley1975_kdtree,
author = {Bentley, Jon Louis},
title = {Multidimensional binary search trees used for associative searching},
year = {1975},
issue_date = {Sept. 1975},
publisher = {Association for Computing Machinery},
address = {New York, NY, USA},
volume = {18},
number = {9},
issn = {0001-0782},
url = {https://doi.org/10.1145/361002.361007},
doi = {10.1145/361002.361007},
journal = {Commun. ACM},
month = sep,
pages = {509–517},
numpages = {9}
}

@ARTICLE{muja2014_flann,
  author={Muja, Marius and Lowe, David G.},
  journal={IEEE Transactions on Pattern Analysis and Machine Intelligence}, 
  title={Scalable Nearest Neighbor Algorithms for High Dimensional Data}, 
  year={2014},
  volume={36},
  number={11},
  pages={2227-2240},
  doi={10.1109/TPAMI.2014.2321376}
}

@inproceedings{guo2020_scann,
author = {Guo, Ruiqi and Sun, Philip and Lindgren, Erik and Geng, Quan and Simcha, David and Chern, Felix and Kumar, Sanjiv},
title = {Accelerating large-scale inference with anisotropic vector quantization},
year = {2020},
publisher = {JMLR.org},
booktitle = {Proceedings of the 37th International Conference on Machine Learning},
articleno = {364},
numpages = {10},
series = {ICML'20}
}

@INPROCEEDINGS{sivic2003_videogoogle,
  author={Sivic and Zisserman},
  booktitle={Proceedings Ninth IEEE International Conference on Computer Vision}, 
  title={Video Google: a text retrieval approach to object matching in videos}, 
  year={2003},
  volume={},
  number={},
  pages={1470-1477 vol.2},
  doi={10.1109/ICCV.2003.1238663}}

@article{jing2024_llm_vecdb,
    title={When Large Language Models Meet Vector Databases: A Survey}, 
    author={Zhi Jing and Yongye Su and Yikun Han and Bo Yuan and Haiyun Xu and Chunjiang Liu and Kehai Chen and Min Zhang},
    year={2024},
    eprint={2402.01763},
    archivePrefix={arXiv},
    primaryClass={cs.DB},
    url={https://arxiv.org/abs/2402.01763}, 
    journal = {CoRR},
    volume = {arXiv:2402.01763},
}

@article{survey2023vdbms1,
    title={A Comprehensive Survey on Vector Database: Storage and Retrieval Technique, Challenge}, 
    author={Yikun Han and Chunjiang Liu and Pengfei Wang},
    year={2023},
    eprint={2310.11703},
    archivePrefix={arXiv},
    primaryClass={cs.DB},
    journal = {CoRR},
    volume = {arXiv:2310.11703},
}

@article{survey2023vdbms2,
author = {Pan, James Jie and Wang, Jianguo and Li, Guoliang},
title = {Survey of vector database management systems},
year = {2024},
issue_date = {Sep 2024},
publisher = {Springer-Verlag},
address = {Berlin, Heidelberg},
volume = {33},
number = {5},
issn = {1066-8888},
url = {https://doi.org/10.1007/s00778-024-00864-x},
doi = {10.1007/s00778-024-00864-x},
journal = {The VLDB Journal},
month = jul,
pages = {1591–1615},
numpages = {25}
}

@inproceedings{wang_milvus_vecdb,
    author = {Wang, Jianguo and Yi, Xiaomeng and Guo, Rentong and Jin, Hai and Xu, Peng and Li, Shengjun and Wang, Xiangyu and Guo, Xiangzhou and Li, Chengming and Xu, Xiaohai and Yu, Kun and Yuan, Yuxing and Zou, Yinghao and Long, Jiquan and Cai, Yudong and Li, Zhenxiang and Zhang, Zhifeng and Mo, Yihua and Gu, Jun and Jiang, Ruiyi and Wei, Yi and Xie, Charles},
    title = {Milvus: A Purpose-Built Vector Data Management System},
    year = {2021},
    isbn = {9781450383431},
    publisher = {Association for Computing Machinery},
    address = {New York, NY, USA},
    url = {https://doi.org/10.1145/3448016.3457550},
    doi = {10.1145/3448016.3457550},
    booktitle = {Proceedings of the 2021 International Conference on Management of Data},
    pages = {2614–2627},
    numpages = {14},
    location = {Virtual Event, China},
    series = {SIGMOD '21}
}

@inproceedings{yang_pase,
    title = {PASE: PostgreSQL Ultra-High-Dimensional Approximate Nearest Neighbor Search Extension},
    author = {Yang, Wen and Li, Tao and Fang, Gai and Wei, Hong},
    year = {2020},
    isbn = {9781450367356},
    publisher = {Association for Computing Machinery},
    address = {New York, NY, USA},
    url = {https://doi.org/10.1145/3318464.3386131},
    doi = {10.1145/3318464.3386131},
    booktitle = {Proceedings of the 2020 ACM SIGMOD International Conference on Management of Data},
    pages = {2241–2253},
    numpages = {13},
    location = {Portland, OR, USA},
    series = {SIGMOD '20}
}

@article{chavez2001_range_search,
author = {Ch\'{a}vez, Edgar and Navarro, Gonzalo and Baeza-Yates, Ricardo and Marroqu\'{\i}n, Jos\'{e} Luis},
title = {Searching in metric spaces},
year = {2001},
issue_date = {September 2001},
publisher = {Association for Computing Machinery},
address = {New York, NY, USA},
volume = {33},
number = {3},
issn = {0360-0300},
url = {https://doi.org/10.1145/502807.502808},
doi = {10.1145/502807.502808},
journal = {ACM Comput. Surv.},
month = sep,
pages = {273–321},
numpages = {49}
}

@article{wu2022_hqann,
    title={HQANN: Efficient and Robust Similarity Search for Hybrid Queries with Structured and Unstructured Constraints}, 
    author={Wei Wu and Junlin He and Yu Qiao and Guoheng Fu and Li Liu and Jin Yu},
    year={2022},
    eprint={2207.07940},
    archivePrefix={arXiv},
    primaryClass={cs.DB},
    url={https://arxiv.org/abs/2207.07940}, 
    journal = {CoRR},
    volume = {arXiv:2207.07940},
}

@article{wang2022_navigable_pg,
    title={Navigable Proximity Graph-Driven Native Hybrid Queries with Structured and Unstructured Constraints}, 
    author={Mengzhao Wang and Lingwei Lv and Xiaoliang Xu and Yuxiang Wang and Qiang Yue and Jiongkang Ni},
    year={2022},
    eprint={2203.13601},
    archivePrefix={arXiv},
    primaryClass={cs.DB},
    url={https://arxiv.org/abs/2203.13601}, 
    journal = {CoRR},
    volume = {arXiv:2203.13601},
}

@article{devlin2019_bert,
      title={BERT: Pre-training of Deep Bidirectional Transformers for Language Understanding}, 
      author={Jacob Devlin and Ming-Wei Chang and Kenton Lee and Kristina Toutanova},
      year={2019},
      eprint={1810.04805},
      archivePrefix={arXiv},
      primaryClass={cs.CL},
      url={https://arxiv.org/abs/1810.04805}, 
      journal = {CoRR},
      volume = {arXiv:1810.04805},
}

@article{li2023_gte_embedding,
      title={Towards General Text Embeddings with Multi-stage Contrastive Learning}, 
      author={Zehan Li and Xin Zhang and Yanzhao Zhang and Dingkun Long and Pengjun Xie and Meishan Zhang},
      year={2023},
      eprint={2308.03281},
      archivePrefix={arXiv},
      primaryClass={cs.CL},
      url={https://arxiv.org/abs/2308.03281}, 
      journal = {CoRR},
      volume = {arXiv:2308.03281},
}

@article{zhang2025_qwen3_embedding,
      title={Qwen3 Embedding: Advancing Text Embedding and Reranking Through Foundation Models}, 
      author={Yanzhao Zhang and Mingxin Li and Dingkun Long and Xin Zhang and Huan Lin and Baosong Yang and Pengjun Xie and An Yang and Dayiheng Liu and Junyang Lin and Fei Huang and Jingren Zhou},
      year={2025},
      eprint={2506.05176},
      archivePrefix={arXiv},
      primaryClass={cs.CL},
      url={https://arxiv.org/abs/2506.05176}, 
      journal = {CoRR},
      volume = {arXiv:2506.05176},
}

@article{openai2023gpt4,
    title={GPT-4 Technical Report}, 
    author={OpenAI},
    year={2023},
    eprint={2303.08774},
    archivePrefix={arXiv},
    primaryClass={cs.CL},
    journal = {CoRR},
    volume = {arXiv:2303.08774},
}

@inproceedings{muennighoff2023_mteb,
    title = "{MTEB}: Massive Text Embedding Benchmark",
    author = "Muennighoff, Niklas  and
      Tazi, Nouamane  and
      Magne, Loic  and
      Reimers, Nils",
    editor = "Vlachos, Andreas  and
      Augenstein, Isabelle",
    booktitle = "Proceedings of the 17th Conference of the European Chapter of the Association for Computational Linguistics",
    month = may,
    year = "2023",
    address = "Dubrovnik, Croatia",
    publisher = "Association for Computational Linguistics",
    url = "https://aclanthology.org/2023.eacl-main.148/",
    doi = "10.18653/v1/2023.eacl-main.148",
    pages = "2014--2037"
}

@article{JOB_benchmark,
author = {Leis, Viktor and Gubichev, Andrey and Mirchev, Atanas and Boncz, Peter and Kemper, Alfons and Neumann, Thomas},
title = {How good are query optimizers, really?},
year = {2015},
issue_date = {November 2015},
publisher = {VLDB Endowment},
volume = {9},
number = {3},
issn = {2150-8097},
url = {https://doi.org/10.14778/2850583.2850594},
doi = {10.14778/2850583.2850594},
journal = {Proc. VLDB Endow.},
month = nov,
pages = {204–215},
numpages = {12}
}

@article{lewis2021_rag,
    title={Retrieval-Augmented Generation for Knowledge-Intensive NLP Tasks}, 
    author={Patrick Lewis and Ethan Perez and Aleksandra Piktus and Fabio Petroni and Vladimir Karpukhin and Naman Goyal and Heinrich Küttler and Mike Lewis and Wen-tau Yih and Tim Rocktäschel and Sebastian Riedel and Douwe Kiela},
    year={2021},
    eprint={2005.11401},
    archivePrefix={arXiv},
    primaryClass={cs.CL},
    url={https://arxiv.org/abs/2005.11401}, 
    journal = {CoRR},
    volume = {arXiv:2005.11401},
}

@inproceedings{thakur2021_beir,
 author = {Thakur, Nandan and Reimers, Nils and R\"{u}ckl\'{e}, Andreas and Srivastava, Abhishek and Gurevych, Iryna},
 booktitle = {Proceedings of the Neural Information Processing Systems Track on Datasets and Benchmarks},
 editor = {J. Vanschoren and S. Yeung},
 pages = {},
 title = {BEIR: A Heterogeneous Benchmark for Zero-shot Evaluation of Information Retrieval Models},
 url = {https://datasets-benchmarks-proceedings.neurips.cc/paper_files/paper/2021/file/65b9eea6e1cc6bb9f0cd2a47751a186f-Paper-round2.pdf},
 volume = {1},
 year = {2021}
}

@article{ji2023_hallucination,
author = {Ji, Ziwei and Lee, Nayeon and Frieske, Rita and Yu, Tiezheng and Su, Dan and Xu, Yan and Ishii, Etsuko and Bang, Ye Jin and Madotto, Andrea and Fung, Pascale},
title = {Survey of Hallucination in Natural Language Generation},
year = {2023},
issue_date = {December 2023},
publisher = {Association for Computing Machinery},
address = {New York, NY, USA},
volume = {55},
number = {12},
issn = {0360-0300},
url = {https://doi.org/10.1145/3571730},
doi = {10.1145/3571730},
journal = {ACM Comput. Surv.},
month = mar,
articleno = {248},
numpages = {38}
}

@inproceedings{covington2016_youtube_recs,
author = {Covington, Paul and Adams, Jay and Sargin, Emre},
title = {Deep Neural Networks for YouTube Recommendations},
year = {2016},
isbn = {9781450340359},
publisher = {Association for Computing Machinery},
address = {New York, NY, USA},
url = {https://doi.org/10.1145/2959100.2959190},
doi = {10.1145/2959100.2959190},
booktitle = {Proceedings of the 10th ACM Conference on Recommender Systems},
pages = {191–198},
numpages = {8},
location = {Boston, Massachusetts, USA},
series = {RecSys '16}
}

@article{analyticDBV-hybrid,
author = {Wei, Chuangxian and Wu, Bin and Wang, Sheng and Lou, Renjie and Zhan, Chaoqun and Li, Feifei and Cai, Yuanzhe},
title = {AnalyticDB-V: a hybrid analytical engine towards query fusion for structured and unstructured data},
year = {2020},
issue_date = {August 2020},
publisher = {VLDB Endowment},
volume = {13},
number = {12},
issn = {2150-8097},
url = {https://doi.org/10.14778/3415478.3415541},
doi = {10.14778/3415478.3415541},
journal = {Proc. VLDB Endow.},
month = aug,
pages = {3152–3165},
numpages = {14}
}

@inproceedings{jd_platform,
author = {Li, Jie and Liu, Haifeng and Gui, Chuanghua and Chen, Jianyu and Ni, Zhenyuan and Wang, Ning and Chen, Yuan},
title = {The Design and Implementation of a Real Time Visual Search System on JD E-commerce Platform},
year = {2018},
isbn = {9781450360166},
publisher = {Association for Computing Machinery},
address = {New York, NY, USA},
url = {https://doi.org/10.1145/3284028.3284030},
doi = {10.1145/3284028.3284030},
booktitle = {Proceedings of the 19th International Middleware Conference Industry},
pages = {9–16},
numpages = {8},
location = {Rennes, France},
series = {Middleware '18}
}

@inproceedings{gpt3,
author = {Brown, Tom B. and Mann, Benjamin and Ryder, Nick and Subbiah, Melanie and Kaplan, Jared and Dhariwal, Prafulla and Neelakantan, Arvind and Shyam, Pranav and Sastry, Girish and Askell, Amanda and Agarwal, Sandhini and Herbert-Voss, Ariel and Krueger, Gretchen and Henighan, Tom and Child, Rewon and Ramesh, Aditya and Ziegler, Daniel M. and Wu, Jeffrey and Winter, Clemens and Hesse, Christopher and Chen, Mark and Sigler, Eric and Litwin, Mateusz and Gray, Scott and Chess, Benjamin and Clark, Jack and Berner, Christopher and McCandlish, Sam and Radford, Alec and Sutskever, Ilya and Amodei, Dario},
title = {Language models are few-shot learners},
year = {2020},
isbn = {9781713829546},
publisher = {Curran Associates Inc.},
address = {Red Hook, NY, USA},
booktitle = {Proceedings of the 34th International Conference on Neural Information Processing Systems},
articleno = {159},
numpages = {25},
location = {Vancouver, BC, Canada},
series = {NIPS '20}
}

@article{parlay_ann2024,
    title={ParlayANN: Scalable and Deterministic Parallel Graph-Based Approximate Nearest Neighbor Search Algorithms},
    author={Magdalen Dobson Manohar and Zheqi Shen and Guy E. Blelloch and Laxman Dhulipala and Yan Gu and Harsha Vardhan Simhadri and Yihan Sun},
    year={2024},
    eprint={2305.04359},
    archivePrefix={arXiv},
    primaryClass={cs.IR},
    url={https://arxiv.org/abs/2305.04359}, 
    journal = {CoRR},
    volume = {arXiv:2305.04359},
}

@article{lsh_survey2021,
    title={A Survey on Locality Sensitive Hashing Algorithms and their Applications}, 
      author={Omid Jafari and Preeti Maurya and Parth Nagarkar and Khandker Mushfiqul Islam and Chidambaram Crushev},
      year={2021},
      eprint={2102.08942},
      archivePrefix={arXiv},
      primaryClass={cs.DB},
      url={https://arxiv.org/abs/2102.08942}, 
      journal = {CoRR},
      volume = {arXiv:2102.08942},
}

@article{patel_lotuslang2024,
      title={Semantic Operators: A Declarative Model for Rich, AI-based Data Processing}, 
      author={Liana Patel and Siddharth Jha and Melissa Pan and Harshit Gupta and Parth Asawa and Carlos Guestrin and Matei Zaharia},
      year={2025},
      eprint={2407.11418},
      archivePrefix={arXiv},
      primaryClass={cs.DB},
      url={https://arxiv.org/abs/2407.11418}, 
      journal = {CoRR},
      volume = {arXiv:2407.11418},
}

@article{patel_acorn2024,
      title={ACORN: Performant and Predicate-Agnostic Search Over Vector Embeddings and Structured Data}, 
      author={Liana Patel and Peter Kraft and Carlos Guestrin and Matei Zaharia},
      year={2024},
      eprint={2403.04871},
      archivePrefix={arXiv},
      primaryClass={cs.IR},
      url={https://arxiv.org/abs/2403.04871}, 
      journal = {CoRR},
      volume = {arXiv:2403.04871},
}

@inproceedings {zhang_vbase,
author = {Qianxi Zhang and Shuotao Xu and Qi Chen and Guoxin Sui and Jiadong Xie and Zhizhen Cai and Yaoqi Chen and Yinxuan He and Yuqing Yang and Fan Yang and Mao Yang and Lidong Zhou},
title = {{VBASE}: Unifying Online Vector Similarity Search and Relational Queries via Relaxed Monotonicity},
booktitle = {17th USENIX Symposium on Operating Systems Design and Implementation (OSDI 23)},
year = {2023},
isbn = {978-1-939133-34-2},
address = {Boston, MA},
pages = {377--395},
url = {https://www.usenix.org/conference/osdi23/presentation/zhang-qianxi},
publisher = {USENIX Association},
month = jul
}

@inproceedings{aggarwal_metric_curse,
author = {Aggarwal, Charu C. and Hinneburg, Alexander and Keim, Daniel A.},
title = {On the Surprising Behavior of Distance Metrics in High Dimensional Spaces},
year = {2001},
isbn = {3540414568},
publisher = {Springer-Verlag},
address = {Berlin, Heidelberg},
booktitle = {Proceedings of the 8th International Conference on Database Theory},
pages = {420–434},
numpages = {15},
series = {ICDT '01}
}

@inproceedings{karpukhin2020-dense,
    title = "Dense Passage Retrieval for Open-Domain Question Answering",
    author = "Karpukhin, Vladimir  and
      Oguz, Barlas  and
      Min, Sewon  and
      Lewis, Patrick  and
      Wu, Ledell  and
      Edunov, Sergey  and
      Chen, Danqi  and
      Yih, Wen-tau",
    editor = "Webber, Bonnie  and
      Cohn, Trevor  and
      He, Yulan  and
      Liu, Yang",
    booktitle = "Proceedings of the 2020 Conference on Empirical Methods in Natural Language Processing (EMNLP)",
    month = nov,
    year = "2020",
    address = "Online",
    publisher = "Association for Computational Linguistics",
    url = "https://aclanthology.org/2020.emnlp-main.550/",
    doi = "10.18653/v1/2020.emnlp-main.550",
    pages = "6769--6781",
}

@inproceedings{mikolov2013_word2vec,
author = {Mikolov, Tomas and Sutskever, Ilya and Chen, Kai and Corrado, Greg and Dean, Jeffrey},
title = {Distributed representations of words and phrases and their compositionality},
year = {2013},
publisher = {Curran Associates Inc.},
address = {Red Hook, NY, USA},
booktitle = {Proceedings of the 27th International Conference on Neural Information Processing Systems - Volume 2},
pages = {3111–3119},
numpages = {9},
location = {Lake Tahoe, Nevada},
series = {NIPS'13}
}

@inproceedings{robertson1994_bm25,
author = {Robertson, S. E. and Walker, S.},
title = {Some simple effective approximations to the 2-Poisson model for probabilistic weighted retrieval},
year = {1994},
isbn = {038719889X},
publisher = {Springer-Verlag},
address = {Berlin, Heidelberg},
booktitle = {Proceedings of the 17th Annual International ACM SIGIR Conference on Research and Development in Information Retrieval},
pages = {232–241},
numpages = {10},
location = {Dublin, Ireland},
series = {SIGIR '94}
}

@inproceedings{jin2026fast,
  title={Fast Vector Search in PostgreSQL: A Decoupled Approach},
  author={Jin, Chenzhe and Zhang, Yunan and Liu, Jiayi and Wang, Jianguo},
  booktitle={Conference on Innovative Data Systems Research (CIDR)},
  year={2026}
}
}

\clearpage
\appendix

\section{ADDITIONAL EXPERIMENTS}\label{sec:appendix}

\begin{figure}[htbp]
\centering
  \includegraphics[width=0.95\textwidth]{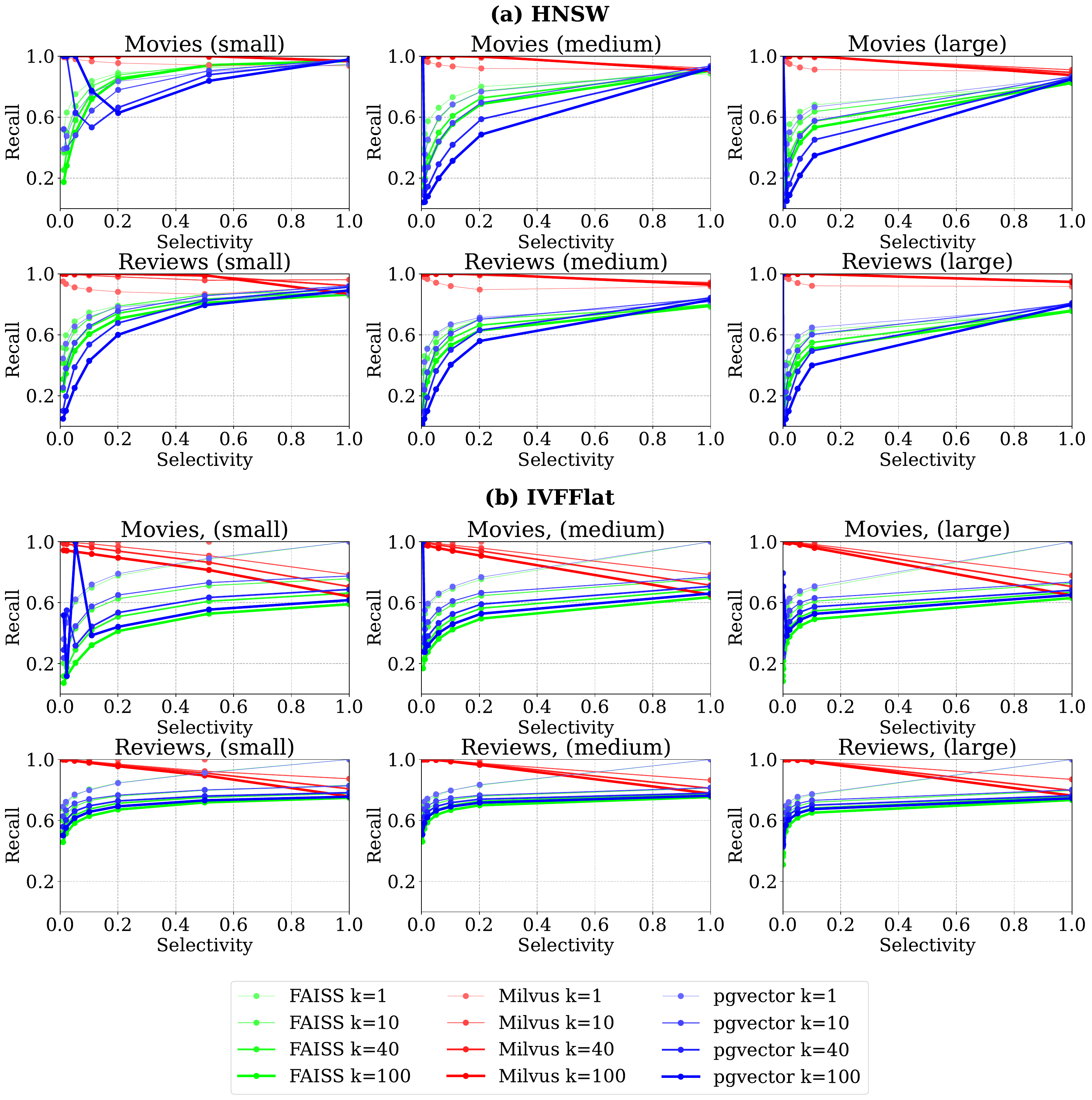}
    \caption{Recall--Selectivity curves for different values of $k$.}
    \label{fig:recall_vs_selectivity_by_k}
\end{figure}

\begin{figure}[htbp]
\centering
  \includegraphics[width=\textwidth]{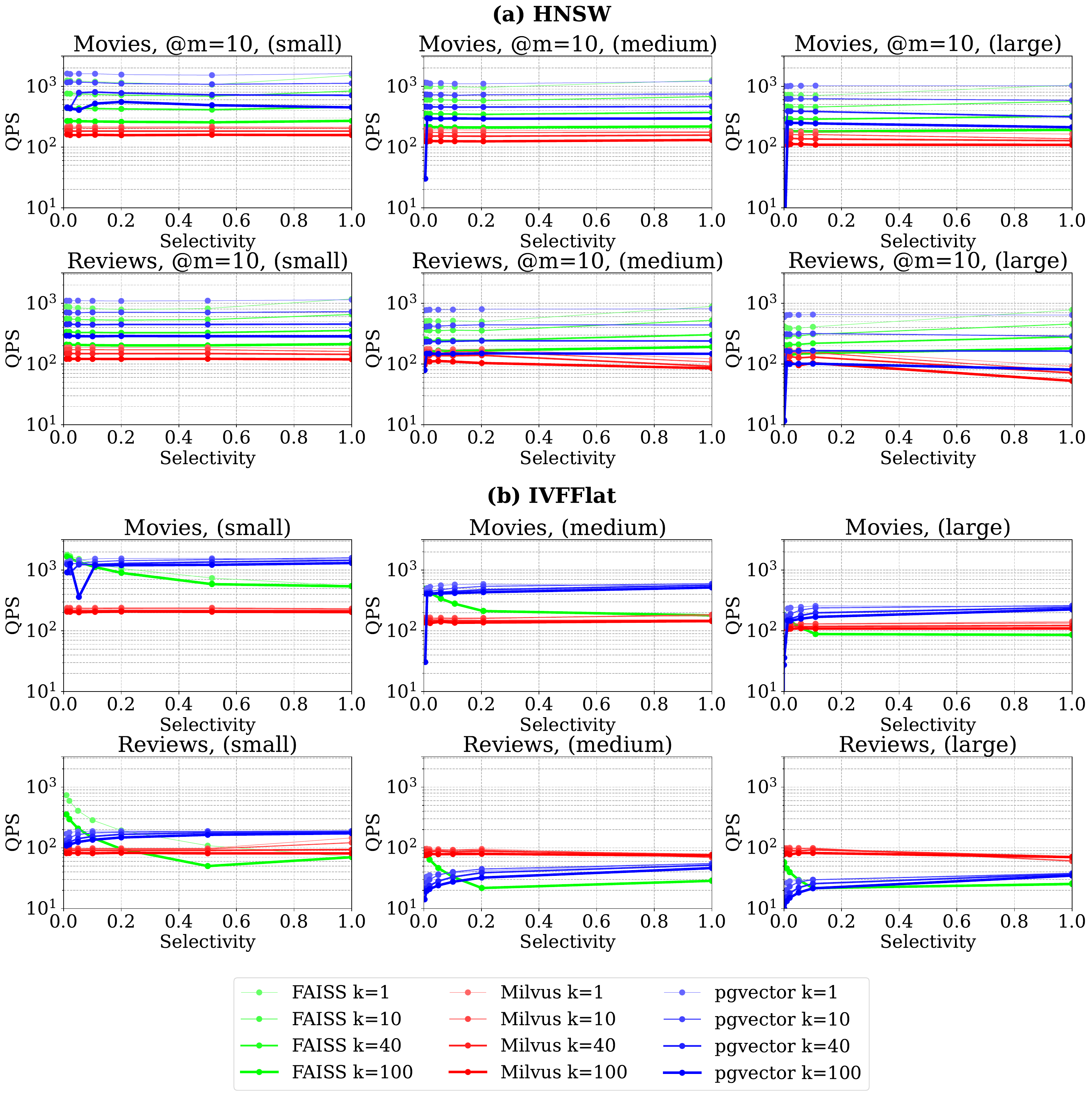}
    \caption{QPS--Selectivity curves for different values of $k$.}
    \label{fig:QPS_sigma_by_k}
\end{figure}

\begin{figure}[htbp]
\centering
  \includegraphics[width=\textwidth]{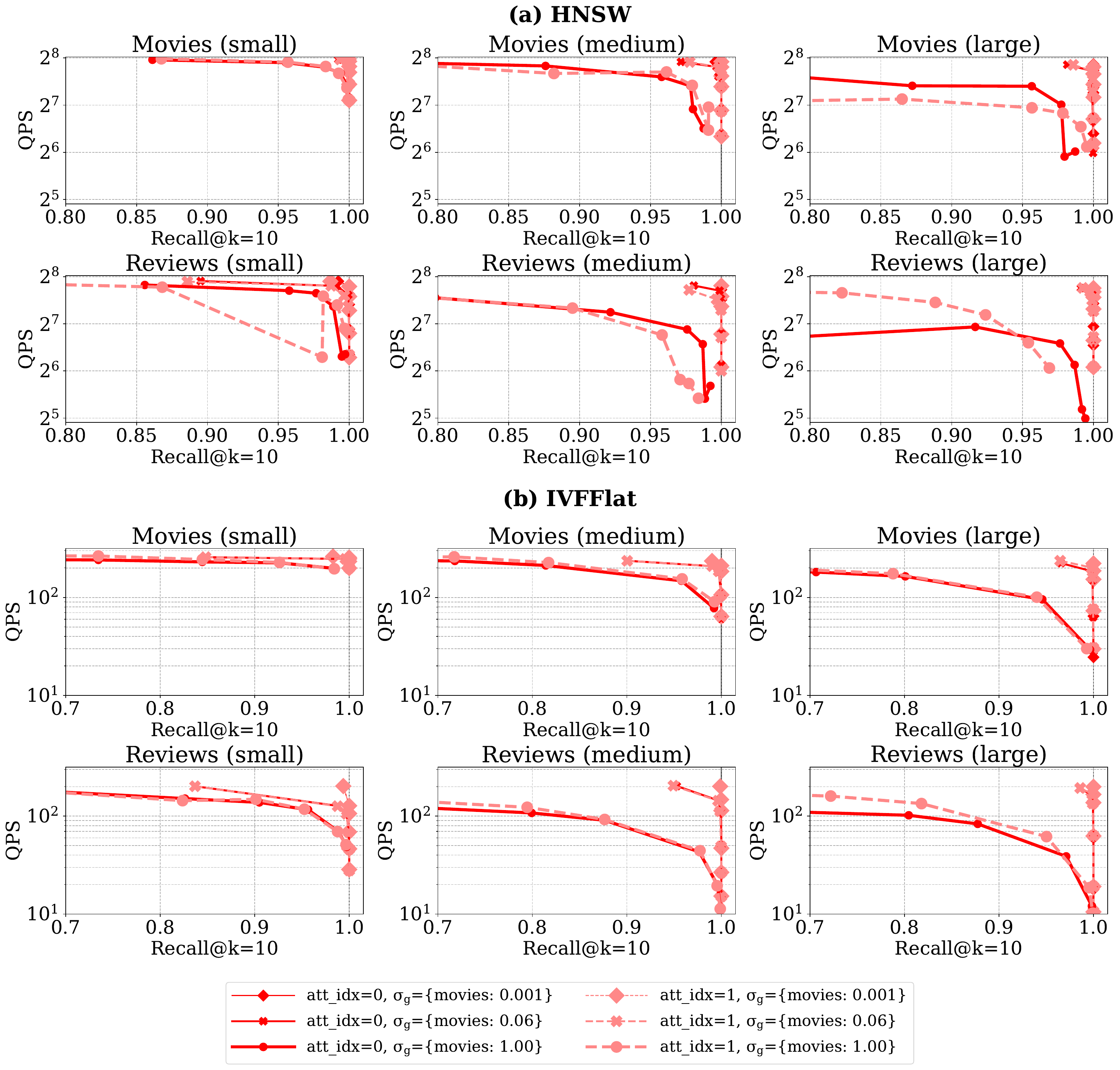}
    \caption{QPS--Recall curves for HNSW and IVFFlat in \textit{Milvus} with and without index on the filter attribute.}
    \label{fig:qps_vs_recall_attidx_milvus}
\end{figure}

\begin{figure}[htbp]
\centering
  \includegraphics[width=\textwidth]{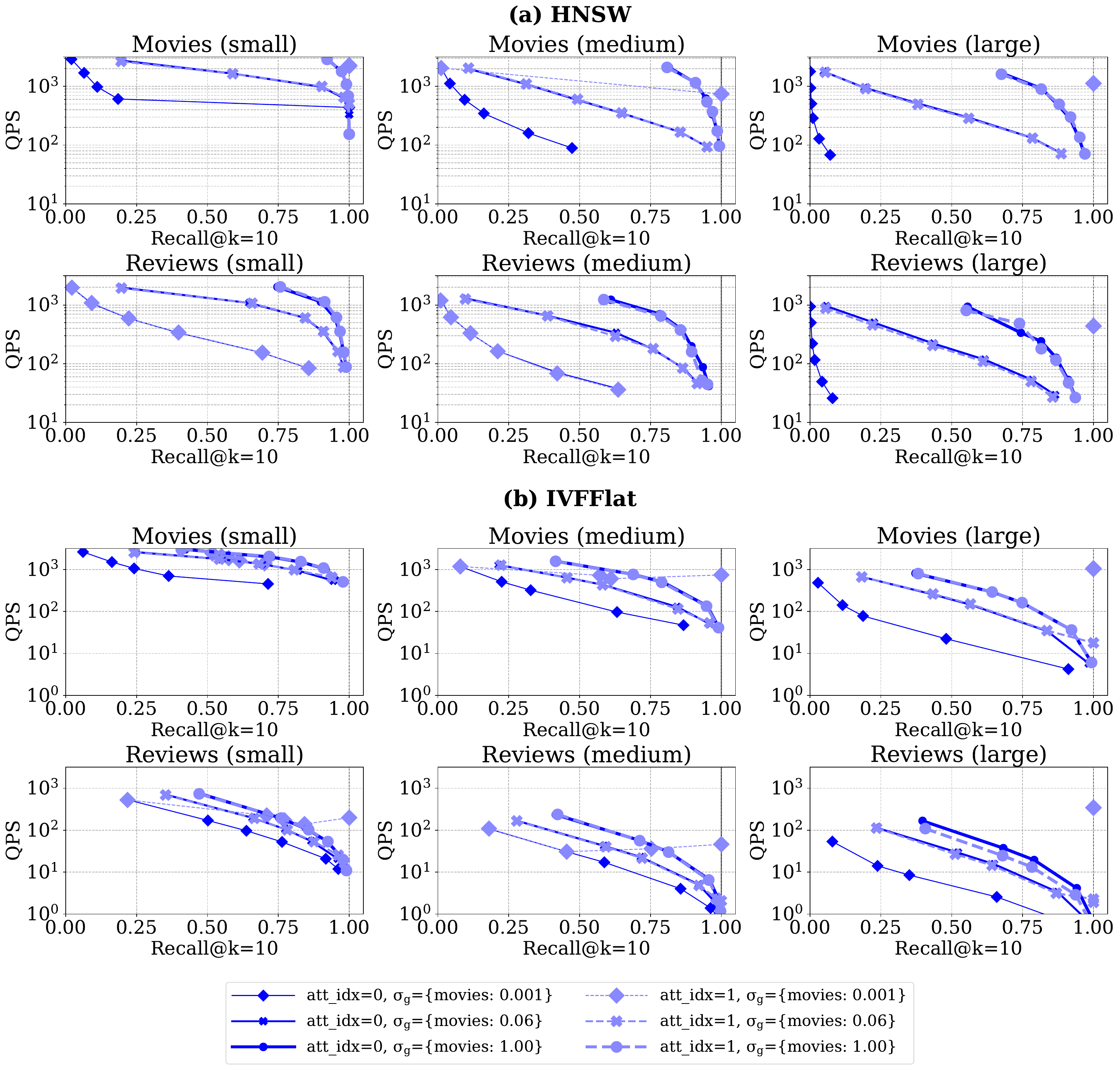}
    \caption{QPS--Recall curves for HNSW and IVFFlat in \textit{pgvector} with and without index on the filter attribute.}
    \label{fig:qps_vs_recall_attidx_pgvector}
\end{figure}


\end{document}